\documentclass   [12pt]  {iopart}

\usepackage  {setspace}
\usepackage  {graphicx, amssymb, amsbsy, color}

\begin  {document}

\title  []
{
Antiferromagnetism and chiral-$d$ wave superconductivity in a honeycomb lattice close to Mott state
}

\author  {Chien-Peng Ho \\ Khee-Kyun Voo*}

\address  {Department of Communication Engineering, Asia Eastern University of Science and Technology, Banciao District, New Taipei City 220303, Taiwan \\
* To whom correspondence should be addressed
(Email: kkvoo@mail.aeust.edu.tw. Tel: +886-2-77388000. Fax: +886-2-77387411)}

%\ead  {kkvoo@mail.oit.edu.tw}

\begin  {abstract}

The antiferromagnetism (AFM) and chiral-$d$ wave superconductivity (SC) in a honeycomb lattice close to an antiferromagnetic (AF) Mott state at half band filling are studied with a $t-J$ model and slave boson mean field theory. The order parameters and single particle dispersion relations at different band filling fractions are investigated. It is found that the AFM enhances the SC, and leads to nodal single particle dispersion relations at two band filling fractions. These unexpected nodal dispersion relations out of nodeless chiral-$d$ wave superconducting order are discussed. A comparison between AF chiral-$d$ wave states and AF extended-$s$ wave states is also given to highlight the pertinent features in chiral-$d$ wave superconducting states.
This study may be related to the honeycomb lattice materials such as the In$_3$Cu$_2$VO$_9$ compound.
\\


\end  {abstract}

\maketitle

\section  {Introduction}
\label  {intro}

%  motivation of studying sc in hubbard honeycomb lattice

Honeycomb lattice materials are interesting for the presence of Dirac single particle dispersion relation \cite {Gei09,NGP09,WKM22}. There are also honeycomb lattice materials such as the In$_3$Cu$_2$VO$_9$ \cite {MLT08,YLZ12} which are insulating and antiferromagnetic (AF) at half band filling (HBF). Given the fact that the composition ions in the material are nonmagnetic, a speculation of the origin of the antiferromagnetism (AFM) and insulating behavior at HBF is a Mott state due to a significant Hubbard repulsion. This might indicate that a strongly correlated system can be obtained when the material is doped with charge carriers.
A lesson from the high temperature superconducting cuprates \cite {PT19} is that such a doped system may exhibit superconductivity (SC), and one can often find an intimate relationship between the Mott state, AFM, and  SC in the system. The SC in the system may be unconventional, since it may not be simply due to exchange of phonons. This motivates us to study the AFM and SC in a honeycomb lattice with a strongly correlated $t-J$ model.

%  lattice struc. and electronic properties (noninteracting).

A two-dimensional honeycomb lattice is a bipartite lattice with two species of sites. The band width is six times the size of the first nearest neighbor (1NN) overlap integral or hopping integral in the lattice. The Fermi surface (FS) of a half filled noninteracting honeycomb lattice consists of two inequivalent Fermi points ($K$ and $K'$) in the Brillouin zone (BZ). The valence and conduction bands touch each other at these two points. The dispersion relation near a Fermi point has the form of a massless Dirac dispersion $\varepsilon_{\bf k} - \mu \propto |{\bf k} - {\bf k} _F |$. Correspondingly, a half filled band has a density of states (DOS) linearly vanishing with energy at approaching the Fermi level (FL). Upon hole doping, disconnected hole-like pieces of FS are opened up from $K$ and $K'$. At 1/4 hole doping density, the two hole-like FSs touch each other at the van Hove singularities, and there is also a FS nesting. Further hole doping merges the two hole-like FSs to a single piece of electron-like FS centered at $\Gamma$.
% a small t' :
% keeps Dirac dispersion cone at FL at HBF?
% keeps vHS at 1/4 hole density?
% FS nesting at 1/4 hole density may lead to a SDW .
% FS nesting is destroyable?

%  expts on honeycomb lattices

The most renown honeycomb lattice material is the graphene or monolayer graphite \cite {Gei09}. 
There are other honeycomb lattice materials like the (111)-bilayer of the perovskite SrIrO$_3$ \cite {XZR11}; the iridates A$_2$IrO$_3$, where A = Na and Li \cite {SMR12,KCK14}; the artificial honeycomb lattice constructed with ultracold atoms \cite {TGU12};
and the insulating and AF compound In$_3$Cu$_2$VO$_9$ \cite {MLT08,YLZ12}.
Superconducting graphenes may be obtained with a proximity effect in a graphene-superconductor heterostructure \cite {LL17},  
chemical doping \cite {LLN15} or electrostatic doping \cite {CFF18,CFD18} a graphene.  

% motivation of t-J model for sc study

In recent years, there has been studies on the SC in doped honeycomb lattices based on the Hubbard and $t-J$ models \cite {SO77,CSO77}.
The effective Hubbard repulsion $U$ in graphene has been estimated by first principle calculation to be $U \simeq 3.3 |t|$ \cite {WSF11}, where $t$ is the 1NN overlap integral in graphene. In a two-dimensional lattice, this $U$ is more than a perturbative interaction strength. On the other hand, the insulating and AF half filled band In$_3$Cu$_2$VO$_9$ \cite {MLT08,YLZ12} might also suggest a considerably strong $U$ in the compound.
The ultra cold atoms honeycomb lattice can also have a relatively strong onsite interatomic repulsion as its band width can be
tuned small. When two layers of graphene are twistedly stacked with a small angle, one can obtain very flat bands at the Fermi level, which may result in a relatively strong interaction between the electrons, and Mott-like insulating states at half-filling can indeed be observed  \cite {CFF18,CFD18}. These observations have motivated the study of SC in the lattice based on strongly correlated models.

% SC theories

The SC in the honeycomb lattice has been investigated in the literature with different models and methodologies.
There were studies with Hubbard type models \cite {PSB10,MHH11,MYY14,QFS17} and $t-J$ type models \cite {GJS13,ZZL15,SH14,Voo20}.
The $t-J$ model \cite {SO77,CSO77} is an effective model for a Hubbard model in the strong $U$ regime, and it operates in a Gutzwiller projected space \cite {Gut63,Vol84} with a suppressed site double occupancy (SDO).
Regarding the methodologies, there were mean field theories \cite {SD07,WSH13,KR86,IMS96,ZGR88,Voo11}, functional renormalization group \cite {WSH13}, quantum Monte Carlo \cite {PSB10,MHH11}, and also Grassman tensor product variational states \cite {GJS13}.
When translational invariant and $C_3$ rotational symmetric superconducting orders are considered \cite {PSB10,MHH11,MYY14,QFS17,GJS13,ZZL15,SH14,SD07,WSH13}, it has been found that a time-reversal asymmetric chiral-$d$ wave SC is the stablest state at close to HBF, whereas an extended-$s$ wave state is the stablest state at band filling fractions below 0.6 \cite {SH14,Voo20,WSH13}. The confluent results in the literatures suggest that the lattice structure and FS geometry are the more decisive factors in the determination of the symmetry of the SC.
When superconducting orders with lower symmetries are also taken into account in a mean field theory (MFT), dimerized states can be found to be stabler at low doping densities \cite {Voo20}. But to what extend the quantum fluctuation underestimated in a MFT may destabilize these dimerized states, and restore the symmetry remains an open question. Therefore a study on the more symmetric chiral-$d$ wave state may still be worthwhile.
Earlier studies considered a singly existing superconducting order in the lattice \cite {PSB10,MHH11,MYY14,QFS17,SH14,SD07,WSH13}, while
a coexisting AFM has also been considered in recent literatures \cite {GJS13,ZZL15,Voo20}.

% paper outline

The study in this paper is based on a $t-J$ model at vanishing SDO and carried out with a slave boson (SB) MFT.
Sec.~\ref {form} introduces the monolayer honeycomb lattice, $t-J$ model, SB MFT formulation, and numerical algorithm. We focus on low hole doping densities.
Sec.~\ref {af-did-op} presents the AF and chiral-$d$ wave superconducting orders in zero and nonzero temperature states.
Sec.~\ref {spx} presents the DOS, single particle dispersion relation, and single particle spectral weight in the AF and chiral-$d$ wave superconducting states.
Sec.~\ref {xs} compares the AF chiral-$d$ wave states with the AF extended-$s$ wave states, to highlight the unique features in the chiral-$d$ wave states. We also discuss the sensitivity of the result to the system parameters.
Sec.~\ref {conc} recollects the finding, compares some results between the $t-J$ models in the honeycomb lattice and the square lattice, and relates the study to the honeycomb lattice materials.

\section  {Model and formulation}
\label  {form}

A $t-J$ model \cite {SO77,CSO77} is the consequence of a strongly repulsive Hubbard model. A strong repulsion between the electrons results in a reduced SDO. Moreover, when an electron virtually hops between two sites with onsite repulsion,  the electron mediates an AF superexchange interaction between the two sites. The superexchange interaction has an exchange integral $J = 4 t^2 / U$, where $t$ and $U$ are the hopping integral and onsite Hubbard repulsion respectively.
A generic $t-J$ model may include one or more electron hopping terms and a Heisenberg exchange interaction term, and the model acts on a Fock space with vanishing SDO. It is taken as a low energy effective model for a repulsive Hubbard model in the strong $U$ limit.

We consider a $t-J$ model for a monolayer honeycomb lattice with a Hamiltonian
\begin {eqnarray}
\hat H = \hat H_t + \hat H_J,
\label {h_tj}
\end {eqnarray}
where $\hat H_t$ and $\hat H_J$ describe the hop of the electrons and Heisenberg superexchange interaction respectively.
We take a SB MFT approach. The SB MFT has been found to reproduce many essential features in a more sophisticated RMFT ~\cite {Voo11}.

A monolayer honeycomb lattice comprises two species of sites. Topologically, a honeycomb lattice is equivalent to a two-orbital square lattice, where an ``orbital'' in the square lattice is a species of sites in the honeycomb lattice. Therefore a honeycomb lattice with $2N$ sites can be formulated by a two-orbital square lattice with $N$ sites.

\begin {figure}

\includegraphics  [scale=0.35]  {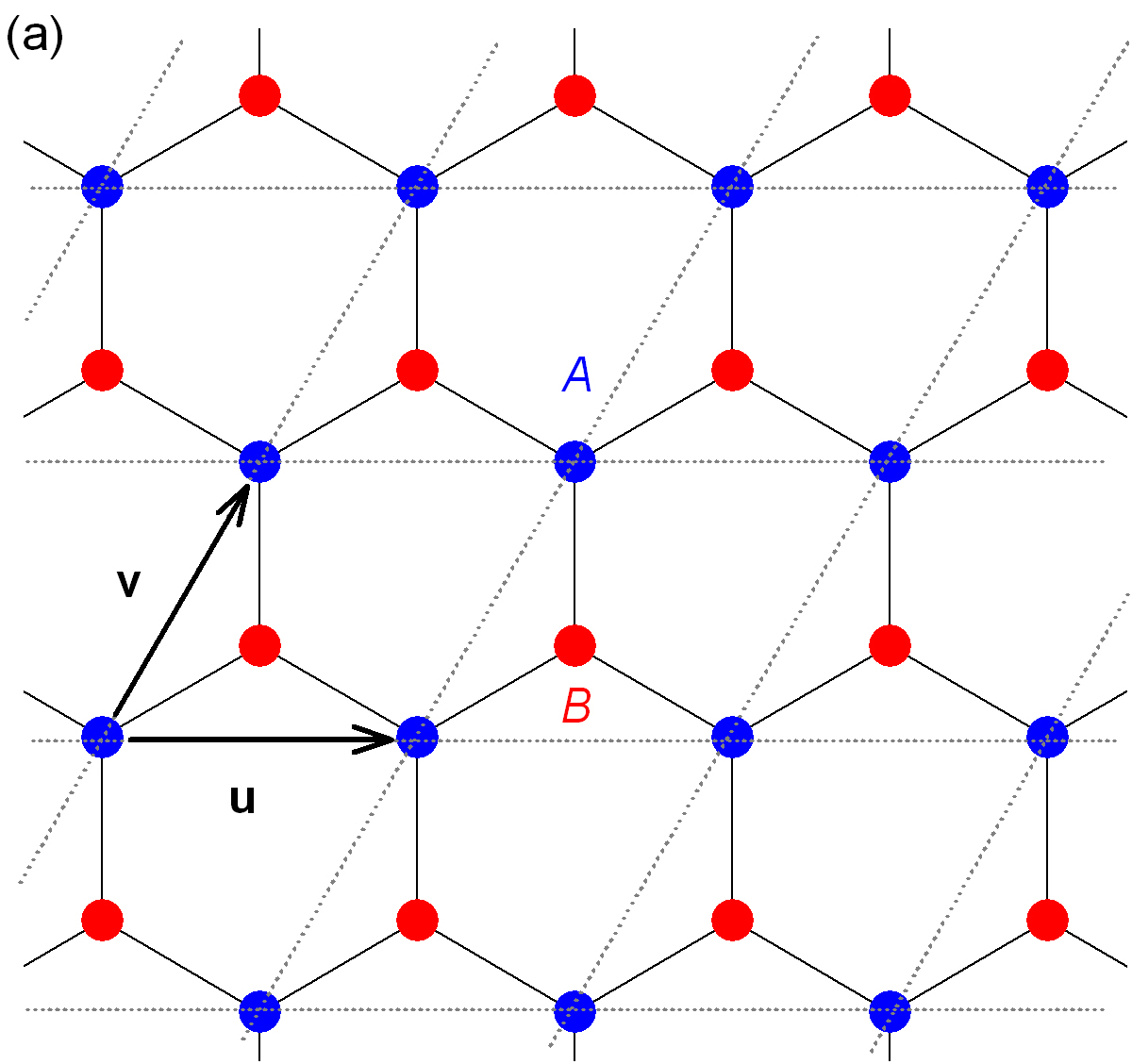} ~~~~~~~~~~~
\includegraphics  [scale=0.33]  {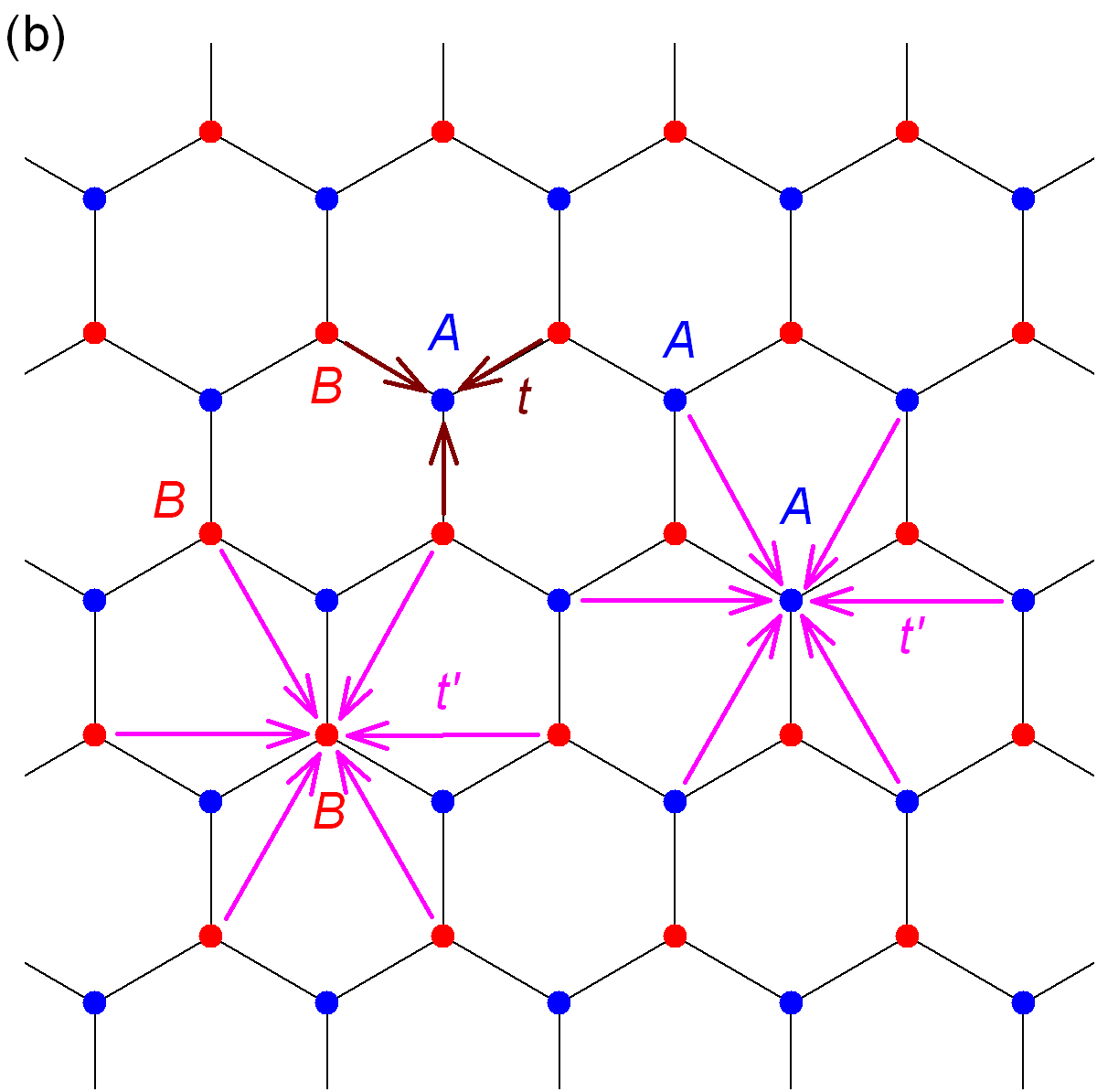} \\
~ \\
\includegraphics  [scale=0.34]  {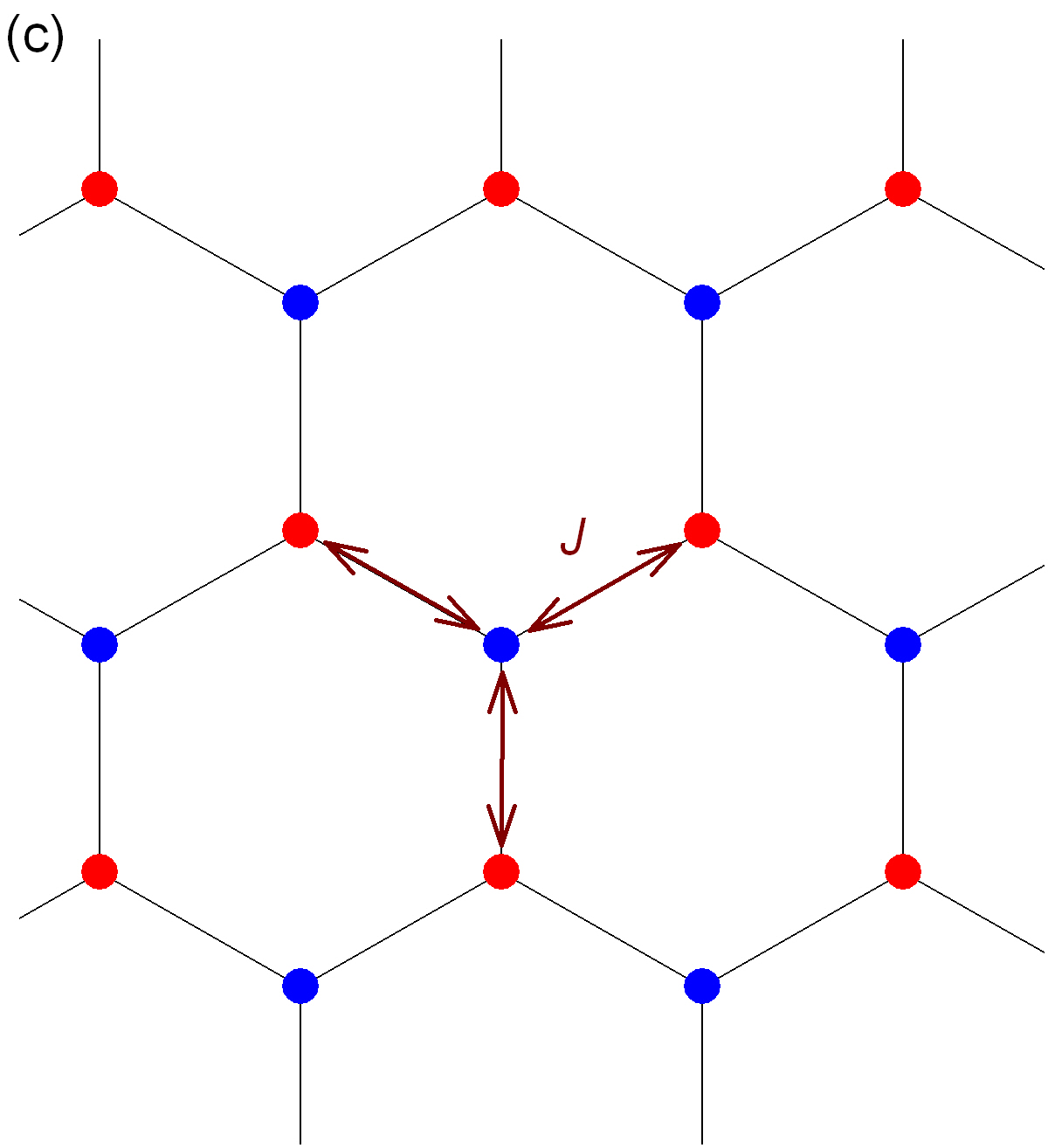} ~~~~~~~~~~~
\includegraphics  [scale=0.34]  {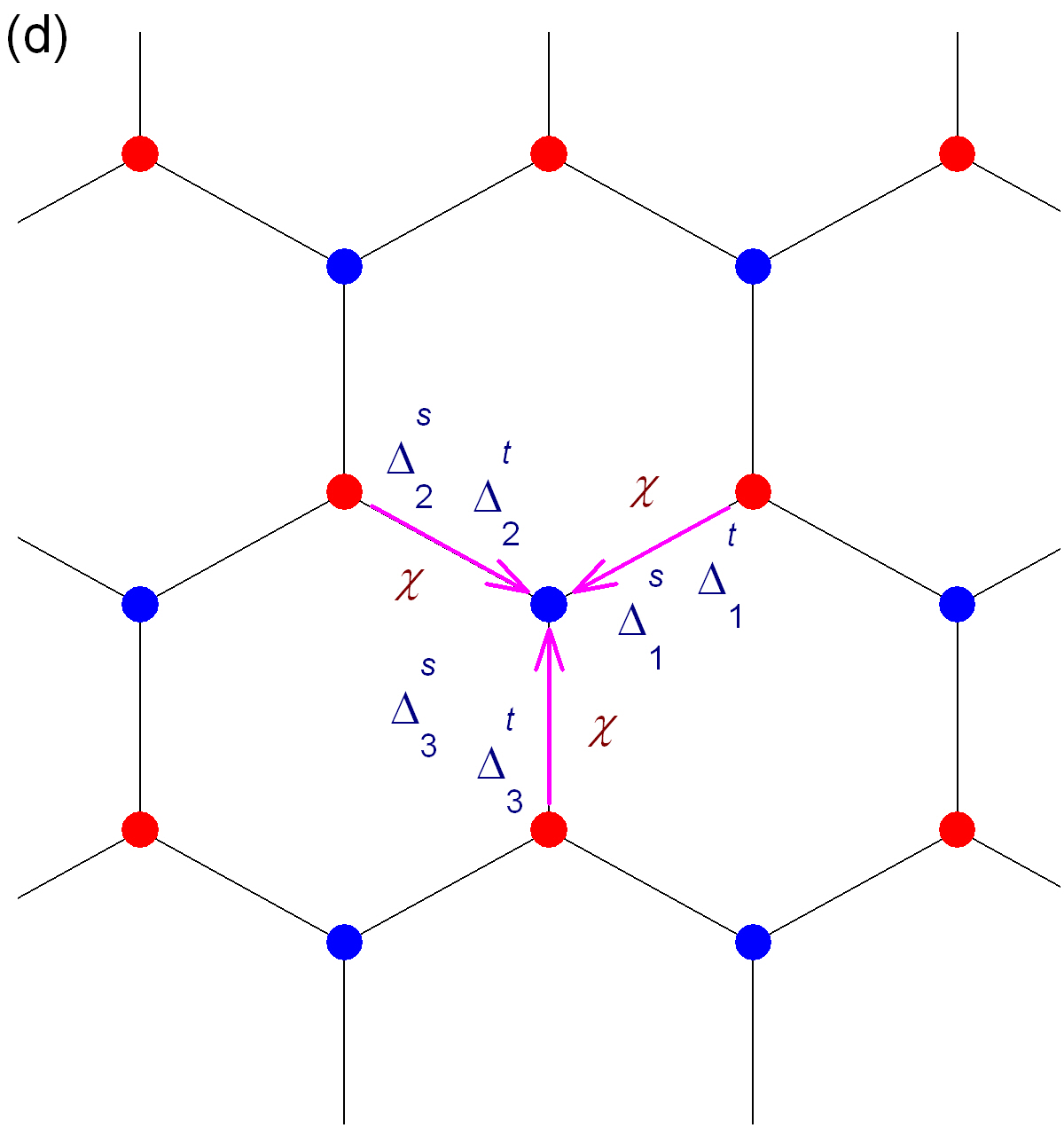} \\

\onehalfspacing

\caption  {
This figure illustrates the honeycomb lattice structure, hopping process, Heisenberg superexchange interaction, and resulting mean field amplitudes.
(a) The backbone sheared square lattice is shown by dotted grid lines. The vectors $\bf u$ and $\bf v$ are the basis vectors.  An unit cell contains an $A$ site (Y-site) and a $B$ site (inverted-Y site).
(b) The 1NN and 2NN hopping with hopping integrals $t$ and $t'$ respectively are depicted.
(c) The 1NN Heisenberg superexchange interaction with an exchange integral $J$ is depicted.
(d) The MF bonding amplitude $\chi$, single pairing amplitudes $\Delta ^s _1$, $\Delta ^s _2$, $\Delta ^s _3$, and  triplet pairing amplitudes $\Delta ^t _1$, $\Delta ^t _2$, $\Delta ^t _3$ are depicted.
}

\label  {lattice}
\end  {figure}

The hopping Hamiltonian is given by
\begin {eqnarray}
\displaystyle
\hat H_t =
\sum_{ {\bf i}, {\bf j}} \sum_ {\alpha, \beta} \sum_ {\sigma }
t _ { {\bf i} \alpha ; {\bf j} \beta }
\hat c ^\dag _ { {\bf i} \alpha \sigma} \hat c ^{~} _ {  {\bf j} \beta \sigma},
\label {h_t}
\end {eqnarray}
where, ${\bf i} = i_u {\bf u} + i_v {\bf v}$ is a lattice site vector, $i_u$ and $i_v$ are integers, and $\bf u$ and $\bf v$ are the lattice basis vectors (likewise is $\bf j$); $\alpha = A$ or $B$ is an orbital index (likewise is $\beta$); and $\sigma = \uparrow$ or $\downarrow$ is a spin index.
Fig.~\ref {lattice}(a) shows a sheared square lattice as the underlying backbone of a honeycomb lattice, a choice of basis vectors ${\bf u} = \hat x$ and ${\bf v} = ( \hat x + \sqrt 3 \hat y ) / 2$; and the $A$ and $B$ orbitals or sites in the honeycomb lattice. We consider isotropic 1NN and second nearest neighbor (2NN) hopping. Fig.~\ref {lattice}(b) shows the interorbital 1NN hopping with integrals $t _ { {\bf i} A ; {\bf i} B } = t _ { {\bf i} A ; {\bf i} - {\bf u}, B } = t _ { {\bf i} A ; {\bf i} - {\bf v}, B } =  t$; and intraorbital 2NN hopping with integrals $t _ { {\bf i} \alpha ; {\bf i} \pm {\bf u}, \alpha } = t _ { {\bf i} \alpha ; {\bf i} \pm {\bf v}, \alpha } = t _ { {\bf i} \alpha ; {\bf i} \pm ({\bf u} - {\bf v}), \alpha } =  t'$, where $\alpha = A$ or $B$.
We let $t = - 1$ and take $|t|$ as the energy unit.

The Heisenberg superexchange interaction Hamiltonian is given by
\begin {eqnarray}
\displaystyle
\hat H_J = {1 \over 2}
\sum_{ {\bf i}, {\bf j}} \sum_ {\alpha, \beta}
J _{{\bf i} \alpha ; {\bf j} {\beta}}
\hat {\bf S} _{ {\bf i} \alpha } \cdot \hat {\bf S} _{ {\bf j} {\beta} },
\label {h_j}
\end {eqnarray}
where $\hat {\bf S} _{ {\bf i} \alpha }$ is an electron spin operator given by $\hat {\bf S} _ {\bf i \alpha} = {\hat S} ^x _{\bf i \alpha} \hat x + {\hat S} ^y _{\bf i \alpha} \hat y + {\hat S} ^z _{\bf i \alpha} \hat z$, ${\hat S} ^ k _{\bf i \alpha} = 2 ^{-1}  \sum _{\mu, \nu} \hat c ^\dagger _{ {\bf i \alpha} \mu} \sigma _k ^{\mu \nu} \hat c ^{~} _{ {\bf i} \alpha \nu}$, and $\sigma _k ^{\mu \nu}$ is the $(\mu,\nu)$ element in the Pauli matrix $\sigma ^{~} _k$ (likewise is $\hat {\bf S} _{ {\bf j} \beta }$). We consider only 1NN interaction, where $J _{{\bf i} \alpha ; {\bf j} {\beta}} = J$ when ${\bf i} \alpha$ and  ${\bf j} {\beta}$ are 1NN of each other, otherwise $J _{{\bf i} \alpha ; {\bf j} {\beta}} = 0$ (see Fig.~\ref {lattice}(c)). This interaction leads to the mean field (MF) bonding amplitude $\chi$, singlet pairing amplitudes $\Delta ^s _1$, $\Delta ^s _2$, $\Delta ^s _3$, and  triplet pairing amplitudes $\Delta ^t _1$, $\Delta ^t _2$, $\Delta ^t _3$ (see Fig.~\ref {lattice}(d)), which will be discussed later in this section.

The $t-J$ Hamiltonian in Eq.~\ref {h_tj} acts in a Fock space in which all doubly occupied sites are excluded. This nonholonomic constraint is difficult to manipulate analytically. For example, it prohibits the Wick's decoupling. Therefore a SB is introduced to reformulate this nonholonomic constraint into a holonomic constraint that can be implemented with a Lagrange multiplier.  First, an electron annihilation operator $\hat c ^{~} _ {  {\bf i} \alpha \sigma}$ is written as the product of a spinless boson creation operator $\hat b ^{\dagger}  _ {  {\bf i} \alpha}$ and a spinful fermion annihilation operator $\hat f  _ {  {\bf i} \alpha \sigma}$, i.e., $\hat c ^{~} _ {  {\bf i} \alpha \sigma} = \hat b ^{\dagger}  _ {  {\bf i} \alpha} \hat f  _ {  {\bf i} \alpha \sigma}$, where $[ \hat b  _ {  {\bf i} \alpha}, \hat b ^{\dagger}  _ {  {\bf j} \beta} ] = \delta _{\bf i \bf j} \delta _{\alpha \beta}$, $\{ \hat f  _ {  {\bf i} \alpha \sigma}, \hat f ^{\dagger}  _ {  {\bf j} \beta \sigma'} \} = \delta _{\bf i \bf j} \delta _{\alpha \beta} \delta _{\sigma \sigma'}$, and $[ \hat b  _ {  {\bf i} \alpha}, \hat b _ {  {\bf j} \beta} ] =  \{ \hat f  _ {  {\bf i} \alpha \sigma}, \hat f  _ {  {\bf j} \beta \sigma'} \} = [ \hat b  _ {  {\bf i} \alpha},  \hat f  _ {  {\bf j} \beta \sigma}    ] = [ \hat b  _ {  {\bf i} \alpha},  \hat f ^ \dagger _ {  {\bf j} \beta \sigma}    ] = 0$. The annihilation of an electron is equivalent to a simultaneous creation of a bosonic hole (holon) and annihilation of a fermionic spin (spinon). The holon is also called a SB. Therefore in a hole doped system, the nonholonomic constraint $\sum _\sigma \hat n _ {  {\bf i} \alpha \sigma}  \leq 1$, where $\hat n _ {  {\bf i} \alpha \sigma}  = \hat c ^{\dagger} _ {  {\bf i} \alpha \sigma} \hat c ^{~} _ {  {\bf i} \alpha \sigma}$, can be rewritten as a holonomic constraint $\hat n _ {  {\bf i} \alpha}^ b + \sum _\sigma \hat n _ {  {\bf i} \alpha \sigma} ^ f = 1$, where $\hat n _ {  {\bf i} \alpha}^ b = \hat b ^{\dagger}  _ {  {\bf i} \alpha} \hat b  _ {  {\bf i} \alpha}  ^{~}$ and $\hat n _ {  {\bf i} \alpha \sigma} ^ f =  \hat f ^{\dagger} _ {  {\bf i} \alpha \sigma} \hat f ^{~} _ {  {\bf i} \alpha \sigma}$. This constraint is treated in a MF manner, where $\langle \hat n _ {  {\bf i} \alpha}^ b \rangle + \sum _\sigma \langle \hat n _ {  {\bf i} \alpha \sigma} ^ f \rangle = 1$.
We consider a charge homogeneous system and we define an average electronic charge density $n =  \sum _\sigma \langle \hat n _ {  {\bf i} \alpha \sigma} \rangle$ and an average hole density
$\delta = 1 - n$.
We let $ \sum _\sigma \langle \hat n _ {  {\bf i} \alpha \sigma} ^ f \rangle  =  n$ and $\langle \hat n _ {  {\bf i} \alpha}^ b \rangle = \delta$.
Moreover, the bosons are assumed to be condensed where a bosonic operator is treated as $\hat b ^{~}  _ {  {\bf i} \alpha} = \hat b ^{\dagger} _ {  {\bf i} \alpha} = \langle \hat b ^{~}  _ {  {\bf i} \alpha} \rangle = \langle \hat b ^{\dagger} _ {  {\bf i} \alpha} \rangle = \sqrt { \delta }$ and $\langle \hat b ^{\dagger}  _ {  {\bf i} \alpha} \hat b  _ {  {\bf i} \alpha} \rangle = \langle \hat b ^{\dagger}  _ {  {\bf i} \alpha} \rangle \langle \hat b  _ {  {\bf i} \alpha} \rangle = \delta$. A hopping term only acts between sites with nonvanishing hole density and it is approximated by $\hat c ^\dag _ { {\bf i} \alpha \sigma} \hat c ^{~} _ {  {\bf j} \beta \sigma} =
 \hat f ^\dagger _ {  {\bf i} \alpha \sigma} \hat b  _ {  {\bf i} \alpha } \cdot \hat b ^{\dagger}  _ {  {\bf j} \beta} \hat f  _ {  {\bf j} \beta \sigma}
= \delta \hat f ^\dag _ { {\bf i} \alpha \sigma} \hat f ^{~} _ {  {\bf j} \beta \sigma}$. The Heisenberg superexchange term acts between singly occupied sites, therefore in a SB formulation it is approximated by a replacement of the electron operator $\hat c  _ {  {\bf i} \alpha \sigma}$ with spinon operator $\hat f  _ {  {\bf i} \alpha \sigma}$.
Hence the $t-J$ Hamiltonian $\hat H$ is approximated by a condensed SB Hamitonian
\begin {eqnarray}
\hat H ^{\rm SB} = \hat H_t ^{\rm SB} + \hat H_J ^{\rm SB},
\label {hsb_tj}
\end {eqnarray}
where
\begin {eqnarray}
\displaystyle
\hat H_t ^{\rm SB} =
\sum_{ {\bf i}, {\bf j}} \sum_ {\alpha, \beta} \sum_ {\sigma }
\delta t _ { {\bf i} \alpha ; {\bf j} \beta } \hat f ^\dag _ { {\bf i} \alpha \sigma} \hat f ^{~} _ {  {\bf j} \beta \sigma},
\label {hsb_t}
\end {eqnarray}
\begin {eqnarray}
\displaystyle
\hat H_J ^{\rm SB} = {1 \over 2}
\sum_{ {\bf i}, {\bf j}} \sum_ {\alpha, \beta}
J _ { {\bf i} \alpha ; {\bf j} \beta }
\hat {\bf S} _{ {\bf i} \alpha } ^{\rm SB} \cdot \hat {\bf S} _{ {\bf j} \beta} ^{\rm SB},
\label {hsb_j}
\end {eqnarray}
where $\hat {\bf S} _{ {\bf i} \alpha } ^{\rm SB}$ is a spinon spin operator given by $\hat {\bf S} _ {\bf i \alpha} ^{\rm SB}= {\hat S} ^ {x, \rm SB} _{\bf i \alpha} \hat x + {\hat S} ^ {y, \rm SB} _{\bf i \alpha} \hat y + {\hat S} ^ {z, \rm SB} _{\bf i \alpha} \hat z$, and ${\hat S} ^ {k, \rm SB} _{\bf i \alpha} = 2^{-1} \sum _{\mu, \nu} \hat f ^\dagger _{ {\bf i \alpha} \mu} \sigma _k ^{\mu \nu} \hat f ^{~} _{ {\bf i} \alpha \nu}$.
Now the Hamiltonian $\hat H ^{\rm SB}$ acts on an ordinary Fock space, where the Wick's decoupling applies. The effect of the exclusion of doubly occupied sites is reflected in the reduction of the band width by a factor of $\delta$.

The Heisenberg interaction term $\hat H_J ^{\rm SB}$ in Eq.~\ref {hsb_j} is further approximated by a quadratic MF Hamiltonian $\hat H_J ^{\rm SB, MF}$, obtained by the following MF decoupling in all channels
\begin  {eqnarray}
\displaystyle
\hat {\bf S} _{ {\bf i} \alpha} ^{\rm SB} \cdot \hat {\bf S} _ { {\bf j} \beta} ^{\rm SB}
&~ \rightarrow ~&
{1 \over 4} \sum _{k = \{ x, y, z \} }  \sum _{\mu, \nu, \gamma, \delta = \{ 1, 2 \}}
\sigma _k ^{\mu \nu} \sigma _k ^{\gamma \delta}  \nonumber \\
&&
\left( ~
\langle \hat f ^\dagger _{ {\bf i} \alpha \mu}  \hat f ^{~} _{ {\bf i} \alpha \nu} \rangle \hat f ^\dagger _{ {\bf j} \beta \gamma}  \hat f ^{~} _{ {\bf j} \beta \delta}
+ \hat  f ^\dagger _{ {\bf i} \alpha \mu}  \hat f ^{~} _{ {\bf i} \alpha \nu} \langle  \hat f ^\dagger _{ {\bf j} \beta \gamma}  \hat f ^{~} _{ {\bf j} \beta \delta} \rangle
- \langle \hat f ^\dagger _{ {\bf i} \alpha \mu}  \hat f ^{~} _{ {\bf i} \alpha \nu} \rangle \langle  \hat f ^\dagger _{ {\bf j} \beta \gamma}  \hat f ^{~} _{ {\bf j} \beta \delta} \rangle \right. \nonumber \\
&&
+ \langle \hat f ^\dagger _{ {\bf i} \alpha \mu} \hat f ^\dagger _{ {\bf j} \beta \gamma}  \rangle \hat   f ^{~} _{ {\bf j} \beta \delta} \hat f ^{~} _{ {\bf i} \alpha \nu}
+  \hat f ^\dagger _{ {\bf i} \alpha \mu}  \hat f ^\dagger _{ {\bf j} \beta \gamma}  \langle \hat   f ^{~} _{ {\bf j} \beta \delta}  \hat f ^{~} _{ {\bf i} \alpha \nu} \rangle
- \langle \hat f ^\dagger _{ {\bf i} \alpha \mu}  \hat f ^\dagger _{ {\bf j} \beta \gamma} \rangle \langle   \hat f ^{~} _{ {\bf j} \beta \delta} \hat f ^{~} _{ {\bf i} \alpha \nu} \rangle \nonumber \\
&&
\left. - \langle \hat f ^\dagger _{ {\bf i} \alpha \mu} \hat  f ^{~} _{ {\bf j} \beta \delta} \rangle \hat f ^\dagger _{ {\bf j} \beta \gamma}  \hat  f ^{~} _{ {\bf i} \alpha \nu}
-  \hat f ^\dagger _{ {\bf i} \alpha \mu} \hat f ^{~} _{ {\bf j} \beta \delta}  \langle \hat  f ^\dagger _{ {\bf j} \beta \gamma}  \hat f ^{~} _{ {\bf i} \alpha \nu} \rangle
+ \langle \hat  f ^\dagger _{ {\bf i} \alpha \mu}  \hat f ^{~} _{ {\bf j} \beta \delta}  \rangle \langle  \hat f ^\dagger _{ {\bf j} \beta \gamma} \hat f ^{~} _{ {\bf i} \alpha \nu}  \rangle \right). \nonumber \\
\label {ss_mf}
\end  {eqnarray}
Hence $\hat H$ is approximated by a quadratic MF Hamiltonian
\begin {eqnarray}
\hat H ^{\rm SB, MF} = \hat H_t ^{\rm SB} + \hat H_J ^{\rm SB, MF},
\label {h_tj_sb_mf}
\end {eqnarray}
in which a condensed boson approximation for the no-double occupancy leading to $\hat H \rightarrow \hat H ^{\rm SB}$ has been made; and a MF decoupling for the superexchange interaction leading to  $\hat H ^{\rm SB} \rightarrow \hat H ^{\rm SB, MF}$ has been made.
A Lagrange multiplier $\mu$ is introduced to impose the constraint for the total number of spinons $ \sum _\sigma \langle \hat n _ {  {\bf i} \alpha \sigma} ^ f \rangle  =  n$, and we solve a grand canonical Hamiltonian
\begin {eqnarray}
\hat K
=\hat H ^ {\rm SB, MF} - \mu \hat N_f,
\label {k}
\end {eqnarray}
where $\hat N _f =  \sum _{ {\bf i} \alpha \sigma} \hat n _ {  {\bf i} \alpha \sigma} ^ f$ is the total number of spinons.

%  Abbreviation and terminology

We define a few abbreviations and terminologies to facilitate the discussion. Throughout this paper, we consider charge homogeneous, translational invariant, and isotropic states. A band filling fraction or site particle density is defined by $n = \sum _\sigma \langle \hat n ^f _ {{\bf i} A \sigma} \rangle = \sum _\sigma \langle \hat n ^f _ {{\bf i} B \sigma} \rangle$. A site hole density is defined by $\delta = 1 - n$.
We consider a collinear AF order with $\langle \hat S ^{x, \rm SB}_{{\bf i} \alpha}  \rangle = \langle \hat S ^ {y, \rm SB} _{{\bf i} \alpha}  \rangle =  0$ and $m = \langle \hat S ^ {z, \rm SB} _{{\bf i} A}  \rangle = - \langle \hat S ^ {z, \rm SB}_{{\bf i} B}  \rangle$.
A real and isotropic 1NN bonding amplitude (BA) is defined by $\chi =  \sum _\sigma  \langle {\hat f} ^\dagger _{{\bf i} A \sigma} {\hat f} ^{} _ {{\bf i}, B, \sigma} \rangle =  \sum _\sigma  \langle {\hat f} ^\dagger _{{\bf i} A \sigma} {\hat f} ^{} _ {{\bf i}-{\hat u}, B, \sigma} \rangle =  \sum _\sigma  \langle {\hat f} ^\dagger _{{\bf i} A \sigma} {\hat f} ^{} _ {{\bf i}-{\hat v}, B, \sigma} \rangle$ (see Fig.~\ref {lattice}(d)).
We define a singlet pairing amplitude (SPA) by
$\Delta ^{ s, {\bf a} }  _ {\alpha \beta} = \langle {\hat f} ^\dagger _{{\bf i} \alpha \uparrow} {\hat f} ^{\dagger} _ {{\bf i}+{\bf a}, \beta, \downarrow} \rangle - \langle {\hat f} ^\dagger _{{\bf i} \alpha \downarrow} {\hat f} ^{\dagger} _ {{\bf i}+{\bf a}, \beta, \uparrow} \rangle$, and a triplet pairing amplitude (TPA) by $\Delta ^{ t, {\bf a} }  _ {\alpha \beta} = \langle {\hat f} ^\dagger _{{\bf i} \alpha \uparrow} {\hat f} ^{\dagger} _ {{\bf i}+{\bf a}, \beta, \downarrow} \rangle + \langle {\hat f} ^\dagger _{{\bf i} \alpha \downarrow} {\hat f} ^{\dagger} _ {{\bf i}+{\bf a}, \beta, \uparrow} \rangle$.
The 1NN SPAs are abbreviated as $\Delta ^s _1 = \Delta ^{s, {\bf 0}} _{AB}$, $\Delta ^s _2 = \Delta ^{s, - {\bf u}} _{AB}$, and $\Delta ^s _3 = \Delta ^{s, - {\bf v}} _{AB}$ (see Fig.~\ref {lattice}(d)); likewise the 1NN TPAs are abbreviated as $\Delta ^t _1 = \Delta ^{t, {\bf 0}} _{AB}$, $\Delta ^t _2 = \Delta ^{t, -{\bf u}} _{AB}$, and $\Delta ^t _3 = \Delta ^{t, -{\bf v}} _{AB}$ (see Fig.~\ref {lattice}(d)).
We consider 1NN pairing amplitudes in the form $\Delta ^s _j = \Delta_s e ^{ i (j-1) \varphi_{\rm sc}}$ and $\Delta ^t _j = \Delta_t e ^{ i (j-1) \varphi_{\rm sc} }$, where $j = 1, 2$, and 3. An isotropic extended-$s$ wave is given by $\varphi_{\rm sc} = 0$.
An isotropic chiral-$d$ wave is given by either $\varphi_{\rm sc} = 2 \pi / 3$ or $\varphi_{\rm sc} = - 2 \pi / 3$ \cite {Voo22pwave}. In other words, chiral-$d$ wave states can be classified into two different types or chiralities by ${\rm sign} (\varphi_{\rm sc})$.
In this paper, we use the notations $d_{x^2-y^2} \pm i d_{xy}$ ($d \pm id'$) for $\varphi_{\rm sc} = \pm 2 \pi / 3$ respectively,
where ``$\pm$" labels the two chiralities of the SC.
The $\Delta_t$ is nonvanishing only when there are coexisting $m$ and $\Delta_s$, since it is due to the inequality of
$| \langle {\hat f} ^\dagger _{{\bf i} \alpha \uparrow} {\hat f} ^{\dagger} _ {{\bf i}+{\bf a}, \beta, \downarrow} \rangle |$ and $| \langle {\hat f} ^\dagger _{{\bf i} \alpha \downarrow} {\hat f} ^{\dagger} _ {{\bf i}+{\bf a}, \beta, \uparrow} \rangle |$ driven by the antiparallel spin moments on the two nearest neighboring sites.
The relative phase between $\Delta_s$ and $\Delta_t$ is tied to the spin orientation. We have ${\rm Arg} ~ (\Delta _s / \Delta_t ) = {\rm Arg} ~ ( m )$, where $m = \langle \hat S ^ {z, \rm SB} _{{\bf i} A}  \rangle$, i.e., when $m > 0$, $\Delta_s$ and $\Delta_t$ are in-phase;  whereas when $m < 0$, $\Delta_s$ and $\Delta_t$ are off-phase by $\pi$.

The translational invariance gives us a convenience to work in a reciprocal space. Defining a spatial Fourier transform for the spinon annihilation operator $\hat f _{ {\bf k} \alpha \sigma} = N ^{- 1/2} \sum _{\bf i} e ^{- i ( k_u i_u + k_v i_v )} \hat f _{ {\bf i} \alpha \sigma}$, where $N$ is the total number of the backbone square lattice sites, and ${\bf k} \equiv ( k_u, k_v )$ is a reciprocal index or wave vector for the site index ${\bf i} \equiv ( i_u, i_v )$, the grand canonical Hamiltonian $\hat K$ can be written as
\begin {eqnarray}
\hat K
&=&
{~~~} \sum_{ {\bf k} \sigma }
\left(  {\hat f} ^\dagger _{{\bf k} A \sigma}  {\hat f} ^{ } _{{\bf k} B \sigma} \varepsilon ^{AB} _{\bf k} + {\rm H. c.} \right)
+
\sum_{ {\bf k} \alpha \sigma }
{\hat f} ^\dagger _{{\bf k} \alpha \sigma}  {\hat f} ^{ } _{{\bf k} \alpha \sigma} \varepsilon ' _{ {\bf k} \alpha \sigma }
+
\sum_{ {\bf k} \sigma }
\left( {\hat f} ^\dagger _{{\bf k} A \sigma}  {\hat f} ^\dagger _{- {\bf k} B \bar \sigma} \Delta^{AB} _{{\bf k} \sigma } + {\rm H. c.} \right)
\nonumber \\
&&
+ E_0,
\label {k_k}
\end {eqnarray}
where
\begin {eqnarray}
\varepsilon^{AB} _{\bf k} = ( \delta t - {3 \over 8} J \chi ) ( 1 + e ^{- i k_u } + e ^ {- i k _v} ), \\
\varepsilon ' _{ {\bf k} \alpha \sigma } = 2 \delta  t' \left[  {\rm cos} ~ (  k_u ) +  {\rm cos} ~ (  k_v  ) +{\rm cos} ~ (  k_u - k_v ) \right]
- {3 \over 2 } \eta_\sigma \eta _\alpha Jm - \mu, \\
\Delta^{AB} _{{\bf k} \sigma} = {1 \over 8} J (-3 \eta_\sigma \Delta_s ^* + \Delta_t ^*) ( 1 + e ^{-i \varphi _{\rm sc} } e ^{- i k_u } + e ^{i \varphi _{\rm sc} } e ^ {- i k _v}), \\
E_0 = 3 N J  m ^ 2 + {9 \over 8} N J \chi ^ 2 + { 3 \over 8 } N J ( 3 |\Delta_s| ^2 - 2 |\Delta_t| ^ 2 ),
\label {}
\end {eqnarray}
and $\eta_\uparrow = \eta_A = 1$ and $\eta_\downarrow = \eta_B = -1$. The magnitude of $\varepsilon^{AB} _{\bf k}$ is maximized at $( k_u, k_v ) = (0,0)$, and nulled at $( k_u, k_v ) = \pm (  - 2 \pi / 3,  2 \pi / 3 )$.
We define $\Gamma = (0,0)$, $K =  ( -  2 \pi / 3,  2 \pi / 3 )$, and $K' = (  2 \pi / 3,  - 2 \pi / 3 )$. Note that $K$ and $K'$ are not equivalent points, and they are related by inversion $( k_u, k_v ) \rightarrow - ( k_u, k_v )$ or interchange $k_u \leftrightarrow k_v$.
In a free band, the single particle dispersion relation has downward Dirac dips and upward Dirac peaks touching each other at $K$ and $K'$.
Note that the $| \Delta^{AB} _{{\bf k} \sigma } |$ for an extended-$s$ wave has a profile identical with that for $| \varepsilon^{AB} _{\bf k} |$, which is maximized at $\Gamma$ and nulled at $K$ and $K'$.
In contrast, a chiral-$d$ wave SC always has $| \Delta^{AB} _{{\bf k} \sigma } | = 0$ at $\Gamma$. When $\varphi _ {\rm sc} = 2 \pi / 3$ ($- 2 \pi / 3$),  $| \Delta^{AB} _{{\bf k} \sigma } |$ is also vanishing at $K'$ ($K$), but is maximized at $K$ ($K'$).
This explains the theoretical finding that an extended-$s$ wave state is favored at near empty bands, but a chiral-$d$ wave is favored at near half filled bands.

The relation between the oblique wave vector $ ( k_u, k_v ) $ and the ordinary rectangular wave vector $ ( k_x,  k_y ) $ is defined by the relation $ k_u i_u + k_v i_v = k_x x + k_y y $, where $ ( x, y ) $ is the rectangular coordinate for a lattice site. The rectangular coordinate $( x, y )$ and oblique coordinate $ ( i_u, i_v ) $ are related by the equality of the expressions of a position vector in the two basis sets, i.e., $x \hat x +  y \hat y = i_u {\bf u} + i_v {\bf v}$. The choice of basis vectors ${\bf u} = \hat x$ and ${\bf v} = ( \hat x + \sqrt 3 \hat y ) / 2$ (Fig.~\ref {lattice}(a)) gives
\begin  {eqnarray}
\left( \begin {array} {c} x \\ y \end {array} \right)
=
\left( \begin {array} {cc}
1 & 1/2 \\ 0 & \sqrt {3} / 2
\end {array} \right)
\left( \begin {array} {c}
i_u \\ i_v
\end {array} \right).
\label {}
\end {eqnarray}
Substituting the above $x$ and $y$ into $k_u i_u + k_v i_v = k_x x + k_y y$, and equating the coefficients of $i_u$ and $i_v$, we obtain the relation between $(k_u,k_v)$ and $(k_x, k_y)$, which is
\begin  {eqnarray}
\left( \begin {array} {c}
k_x \\ k_y
\end {array} \right)
=
\left( \begin {array} {cc}
1 & 0 \\ -1/\sqrt{3} & 2 / \sqrt {3}
\end {array} \right)
\left( \begin {array} {c}
k_u \\ k_v
\end {array} \right)
\label {}
\end {eqnarray}
or
\begin  {eqnarray}
\left( \begin {array} {c}
k_u \\ k_v
\end {array} \right)
=
\left( \begin {array} {cc}
1 & 0 \\ 1/2 & \sqrt {3} / 2
\end {array} \right)
\left( \begin {array} {c}
k_x \\ k_y
\end {array} \right).
\label {}
\end {eqnarray}
The $K$ and $K'$ points at $( k_u, k_v ) =  \eta ( - 2 \pi / 3, 2 \pi / 3 ) + ( 2 m \pi, 2 n \pi )$, where $\eta = 1$ and $-1$ correspond to $K$ and $K'$ respectively; and $m$ and $n$ are integers; are mapped to $( k_x, k_y ) = ( ( - \eta + 3 m ) 2 \pi / 3 ,  ( \eta - m + 2 n ) 2 \pi / \sqrt {3} )$. Discussion of the dispersion relation is more conveniently carried out in the original $( k_u, k_v )$ space.

Discussion in this paper is based on the Matsubara Green function formalism. We define normal and anomalous Matsubara Green functions by $G ^{-+} ( {\bf k} ; \alpha \sigma ,  \beta \sigma' ; \tau) = - \langle T_\tau  \hat f ^{~} _{ {\bf k} \alpha \sigma}  (\tau) \hat f ^ \dagger _{ {\bf k} \beta \sigma'} \rangle$ and $G ^{++} ( {\bf k} ; \alpha \sigma ,  \beta \sigma' ; \tau) = - \langle T_\tau  \hat f ^ \dagger _{ {- \bf k}, \alpha, \sigma}  (\tau) \hat f ^ \dagger _{ {\bf k} \beta \sigma'} \rangle$ respectively.
Since $\hat K$ is time independent, we can perform a temporal Fourier transform to the imaginary frequency space by $G ( i \omega_n) = \int _0 ^\beta G (\tau) e ^{i \omega_n \tau} d \tau$, where $\omega_n$ is a fermionic Matsubara frequency. The normal and anomalous Matsubara Green functions in frequency space are $G ^ {-+} ( {\bf k} ; \alpha \sigma ,  \beta \sigma' ; i \omega_n) $ and $G ^ {++} ( {\bf k} ; \alpha \sigma ,  \beta \sigma' ; i \omega_n) $ respectively. The pairing potential in $\hat H ^{\rm SB, MF} _J$ mixes $G ^ {-+}$ and $G ^ {++}$.
We define $2 \times 2$ Matsubara Green function matrices $\tilde G ^{-+} ( {\bf k, \sigma};  i \omega_n ) $ and $\tilde G ^{++} ( {\bf k, \sigma};  i \omega_n ) $, by $[ \tilde G ^{-+} ( {\bf k, \sigma};  i \omega_n ) ] _ {\zeta_\alpha  \zeta_\beta} =  G ^{-+} ( {\bf k} ; \alpha \sigma ,  \beta \sigma ; i \omega_n)$ and $[ \tilde G ^{++} ( {\bf k, \sigma};  i \omega_n ) ] _ {\zeta_\alpha \zeta_ \beta} =  G ^{++} ( {\bf k} ; \alpha \bar \sigma ,  \beta \sigma ; i \omega_n) $, where $\bar \sigma$ is the opposite spin of $\sigma$,
$\zeta_A=1$, and $\zeta_B=2$. Then the Dyson equations set for the Matsubara Green functions can be compactly written as
\begin {eqnarray}
i \omega_n
\left(  \begin {array} {c}
\tilde G ^{-+} ( {\bf k, \sigma};  i \omega_n ) \\
\tilde G ^{++} ( {\bf k, \sigma};  i \omega_n )
\end {array} \right)
=
\left( \begin {array} {c}
I_2 \\ O_2
\end {array} \right)
+
h _ {\bf k, \sigma}
\left(
\begin {array} {c}
\tilde G ^{-+} ( {\bf k, \sigma};  i \omega_n ) \\
\tilde G ^{++} ( {\bf k, \sigma};  i \omega_n ),
\end {array}
\right),
\label {dyson}
\end {eqnarray}
where $I_2$ is a $2 \times 2$ unit matrix, $O_2$ is a $2 \times 2$ zero matrix, and $h _ {\bf k, \sigma}$ is an interaction matrix given by
\begin {eqnarray}
h _ {\bf k, \sigma} = { h _ {\bf k, \sigma} } ^ \dagger =
\left( \begin {array} {cc}
\tilde V _ {\bf k, \sigma} & \tilde \Delta _ {\bf k, \sigma} \\
( \tilde \Delta _ {\bf k, \sigma} ) ^ \dagger & - \tilde V _ {\bf k, \bar \sigma}
\end {array} \right),
\label {}
\end {eqnarray}
where
\begin {eqnarray}
\tilde V _ {\bf k, \sigma} = { \tilde V _ {\bf k, \sigma} } ^ \dagger =
\left( \begin {array} {cc}
\varepsilon ' _{ {\bf k} A \sigma }
& \varepsilon ^{AB} _{\bf k} \\
( \varepsilon ^{AB} _{\bf k} ) ^*
& \varepsilon ' _{ {\bf k} A \bar \sigma }
\end {array} \right)
\label {}
\end {eqnarray}
and
\begin {eqnarray}
\Delta _ {\bf k, \sigma} =
\left(  \begin {array} {cc}
0 & \Delta ^{AB} _{{\bf k} , \sigma } \\
- \Delta ^{AB} _{- {\bf k} , { \bar \sigma } } & 0
\end {array} \right).
\label {}
\end {eqnarray}
The Green functions can be solved from the Dyson matrix equation (Eq.~\ref {dyson}) by a matrix inversion as
\begin {eqnarray}
\left(  \begin {array} {c}
\tilde G ^{-+} ( {\bf k, \sigma};  i \omega_n ) \\
\tilde G ^{++} ( {\bf k, \sigma};  i \omega_n )
\end {array} \right)
=
{ 1 \over { i \omega_n -  h _ {{\bf k}, \sigma} } }
\left( \begin {array} {c}
I_2 \\ O_2
\end {array} \right).
\label {green-solved}
\end {eqnarray}
Our discussion proceeds by diagonalizing $h _ {{\bf k}, \sigma}$.

The single particle dispersion relation $\varepsilon _{{\bf k},\sigma}$ is found by diagonalizing $h _ {{\bf k}, \sigma}$. The secular equation ${\rm det} ~ ( \varepsilon  _{{\bf k},\sigma} - h _ {{\bf k}, \sigma} ) = 0$ is a quartic equation
\begin {eqnarray}
\varepsilon  _{\bf k, \sigma} ^ 4
-  \varepsilon _{{\bf k},\sigma} ^ 2  B_{{\bf k},\sigma}
+ D_{{\bf k},\sigma} = 0,
\label {secular-eq}
\end {eqnarray}
where
\begin {eqnarray}
B_{{\bf k},\sigma} =  ( \varepsilon ' _{{ \bf k} A \sigma } ) ^ 2
+ ( \varepsilon ' _{{ \bf k} A \bar \sigma } ) ^ 2
+ 2 |  { \varepsilon ^{AB} _{\bf k} }  | ^ 2
+ |  \Delta^{AB} _{{\bf k} , \sigma } | ^ 2
+ | \Delta ^{AB} _{- {\bf k} , { \bar \sigma } } | ^ 2,
\end {eqnarray}
and $D_{{\bf k},\sigma} = {\rm det} ~ ( h _ {\bf k, \sigma} )$, is the determinant of $h _ {\bf k, \sigma}$, given by
\begin {eqnarray}
D_{{\bf k},\sigma}
&=&
( {\varepsilon ' _{ {\bf k} A \sigma } }
{\varepsilon ' _{{ \bf k} A \bar \sigma } } ) ^ 2
- 2  {\varepsilon ' _{ {\bf k} A \sigma } }
{\varepsilon ' _{{ \bf k} A \bar \sigma } } | \varepsilon ^{AB} _{\bf k} | ^ 2
+ | ( \varepsilon ^{AB} _{\bf k} ) ^ 2 -  \Delta^{AB} _{{\bf k} , \sigma } ( \Delta ^{AB} _{- {\bf k} , { \bar \sigma } } ) ^ * | ^ 2
\nonumber  \\
&&
+ | {\varepsilon ' _{ {\bf k} A \sigma } } \Delta ^{AB} _{- {\bf k} , { \bar \sigma } } | ^ 2
+ | {\varepsilon ' _{ {\bf k} A \bar \sigma } }  \Delta^{AB} _{{\bf k} , \sigma } | ^ 2.
\label {det}
\end {eqnarray}
Eq.~\ref {secular-eq} is quadratic in $\varepsilon _{{\bf k},\sigma} ^ 2$ and can be readily solved to give four branches of positive-negative symmetric $\varepsilon _{ {\bf k}, \sigma}$. We denote a dispersion branch as $\varepsilon  _{ {\bf k}, \sigma , j}$, where $j = 1$ to 4, from lowest to highest frequency respectively. The determinant $D_{{\bf k},\sigma} = \prod _{j=1} ^ 4 \varepsilon  _{ {\bf k}, \sigma , j} = ( \prod _{j=1} ^ 2 \varepsilon  _{ {\bf k}, \sigma , j} ) ^2$ is positive definite. Closing of gap or a zero frequency solution occurs if and only if $D_{{\bf k},\sigma} = 0$.

% spectral weight and dos

A spin resolved single particle spectral weight function is defined by $A ( {\bf k}, \sigma; \omega ) = - \pi ^ {-1} \sum _ {\alpha}  {\rm Im} ~ [ G ^{-+} ( {\bf k} ; \alpha \sigma ,  \alpha \sigma ; i \omega_n )  ] _{i \omega _n \rightarrow \omega + i0}$.
The spectral weight function has the form $A ( {\bf k}, \sigma; \omega ) = \sum _{ j = 1} ^ 4 a _{{\bf k}, \sigma, j} \delta ( \omega - \varepsilon _{{\bf k}, \sigma, j} )$, where  $\varepsilon _{{\bf k}, \sigma, j}$ is a dispersion branch, and $a _{{\bf k}, \sigma, j}$ is its branching weight. The branching weights are derived from the eigenvectors of $h _ {{\bf k}, \sigma}$.
The branching weights are positive definite and they respect the  unitarity sum rule $ \sum _{ j = 1} ^ 4 a _{{\bf k}, \sigma, j} = 2$.
The DOS is given by $D (\omega) = N ^ {-1} \sum _{{\bf k},\sigma}  A ({\bf k}, \sigma; \omega)$, which is integrated over an unit BZ where $N$ is the number of integration points.

% symmetry in dispersion

The single particle dispersion relation $\varepsilon _{ {\bf k}, \sigma}$ in an AF chiral-$d$ wave superconducting state depends on the particle spin $\sigma$, site spin moment $m$, and the chirality of the superconducting order ${\rm sign} (\varphi_{\rm sc})$. But there exists a symmetry between the dispersion relations for those degenerated states at the same band filling fraction. First we note that those states have the same $\chi$, $|m|$, $\mu$, $|\Delta_s|$, $|\Delta_t|$, and $|\varphi_{\rm sc}|$.
Then we consider the the $B_{{\bf k},\sigma}$ and $D_{{\bf k},\sigma}$ in the secular equation (Eq.~\ref {secular-eq}).
When we look into $D_{{\bf k},\sigma}$, we see that when we replace $\eta_\sigma m$ with $- \eta_\sigma m$, we get a result that is identical with that from either replacing $\varphi_{\rm sc}$ with $- \varphi_{\rm sc}$, replacing $( k_u, k_v )$ with $(- k_u,- k_v )$, or replacing $(k_u, k_v)$ with $(k_v, k_u)$. Succinctly, $D_{{\bf k},\sigma} | _{\eta_\sigma m \rightarrow - \eta_\sigma m} = D_{{\bf k},\sigma} | _{\varphi_{\rm sc} \rightarrow - \varphi_{\rm sc}} = D_{{\bf k},\sigma} | _{k_u \rightarrow - k_u, ~ k_v \rightarrow - k_v} = D_{{\bf k},\sigma} | _{ k_u \rightarrow k_v, ~ k_v \rightarrow k_u}$.
We also find the same symmetry in $B_{{\bf k},\sigma}$. This leads to the same symmetry in the dispersion relation $\varepsilon _{{\bf k},\sigma}$.
In a nutshell, for those degenerated AF chiral-$d$ wave superconducting states at the same band filling fraction, there are only two distinct dispersion relations, determined by the sign of $\varphi_{\rm sc} \eta_\sigma m$, and the two dispersion relations are related by the symmetry transformation described below.
Let $\varepsilon  _{\bf k} ^ +$ ($\varepsilon  _{\bf k} ^ -$) be the solution of $\varepsilon  _{\bf k, \sigma}$ at $\varphi_{\rm sc} \eta_\sigma m > 0$ ($\varphi_{\rm sc} \eta_\sigma m < 0$), then
an inversion $ ( k_u, k_v ) \rightarrow (- k_u,- k_v ) $ or interchange $(k_u, k_v) \rightarrow (k_v, k_u)$ transforms $\varepsilon  _{\bf k} ^ +$ to $\varepsilon  _{\bf k} ^ -$ or vice versa \cite {Voo22symm}.
An example will be presented in Fig.~\ref {aksw_zoom}(d).
For an extended-$s$ wave state where $\varphi_{\rm sc} = 0$, the secular equation is independent of the sign of $\eta_\sigma m$, and also the solution of $\varepsilon _{{\bf k},\sigma}$. The dispersion relation in this case is symmetric in $ ( k_u, k_v ) \rightarrow (- k_u,- k_v ) $ and $(k_u, k_v) \rightarrow (k_v, k_u)$. A few examples will be presented in Figs.~\ref {xs-ep}(b), \ref {xs-ep}(d), and \ref {xs-ep}(f).

% iteration procedure

The order parameters can be found by a self-consistent numerical iteration on a periodic $\sqrt N \times \sqrt N$ square lattice.
For a given band filling fraction $n$ and temperature $T$,
the iteration begins with a $\hat K$ where the order parameters are assigned with random but reasonable complex values, and the Lagrange multiplier $\mu$ is assigned with a random but reasonable real value. The particle number expectation values are intrinsically real, while the other order parameters may be complex.
New order parameters are evaluated as equal-time Matsubara Green functions, and each order parameter is slowly revised by a weighted average between the old and new values for numerical stability. The Lagrange multiplier $\mu$ is also slowly tuned to produce the desired fermion particle number per honeycomb lattice site $\langle \hat N _f \rangle / (2N) = n$. Hence the grand canonical Hamiltonian $\hat K$ is revised, and we can proceed to the evaluation of new order parameters.
Conditions may be imposed on the order parameters in the iteration (see discussions in later sections).
The evaluation and revision of the order parameters are repeated until an acceptable self-consistency.

We will focus on the low hole doping and AF regime, with a coexisting SC. Nonmagnetic and extended-$s$ wave states will also be mentioned for comparison.
We will mostly consider the system parameters $(t,t';J) = (-1,0;0.5)$, but we will also discuss the robustness of the result with respect to the variation of the system parameters.

\section  {Order parameters in antiferromagnetic chiral-$d$ wave states}
\label  {af-did-op}

We begin with a study on the dependence of the order parameters on the band filling fraction $n$ and temperature $T$. Throughout this section, we consider a system with $(t,t';J) = (-1,0;0.5)$. The AF order parameter $m$ and superconducting order parameter $\Delta_s$ are solved with two sets of conditions. We solve the order parameters with an AF chiral-$d$ wave condition and a nonmagnetic chiral-$d$ wave condition.

% T = 0  plot

\begin {figure}

\includegraphics  [scale=0.38]  {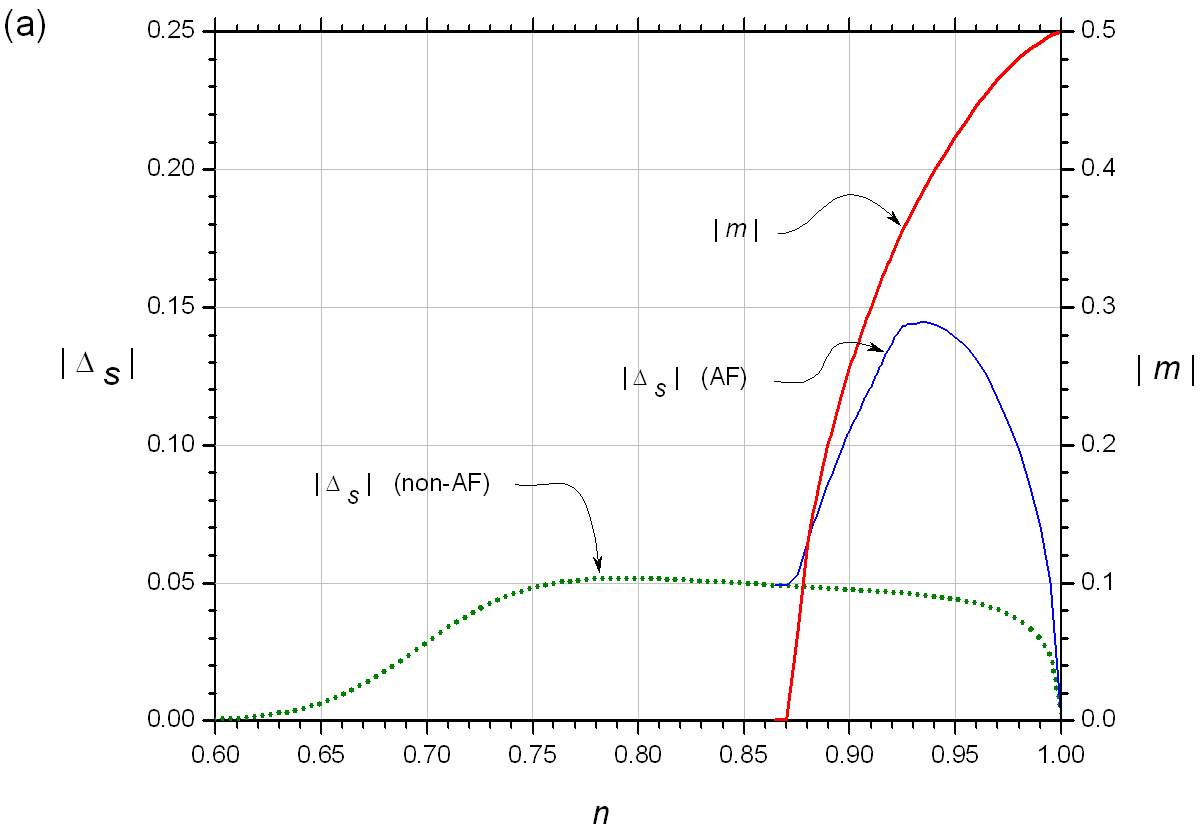} ~~~~~~~~~
\includegraphics  [scale=0.42]  {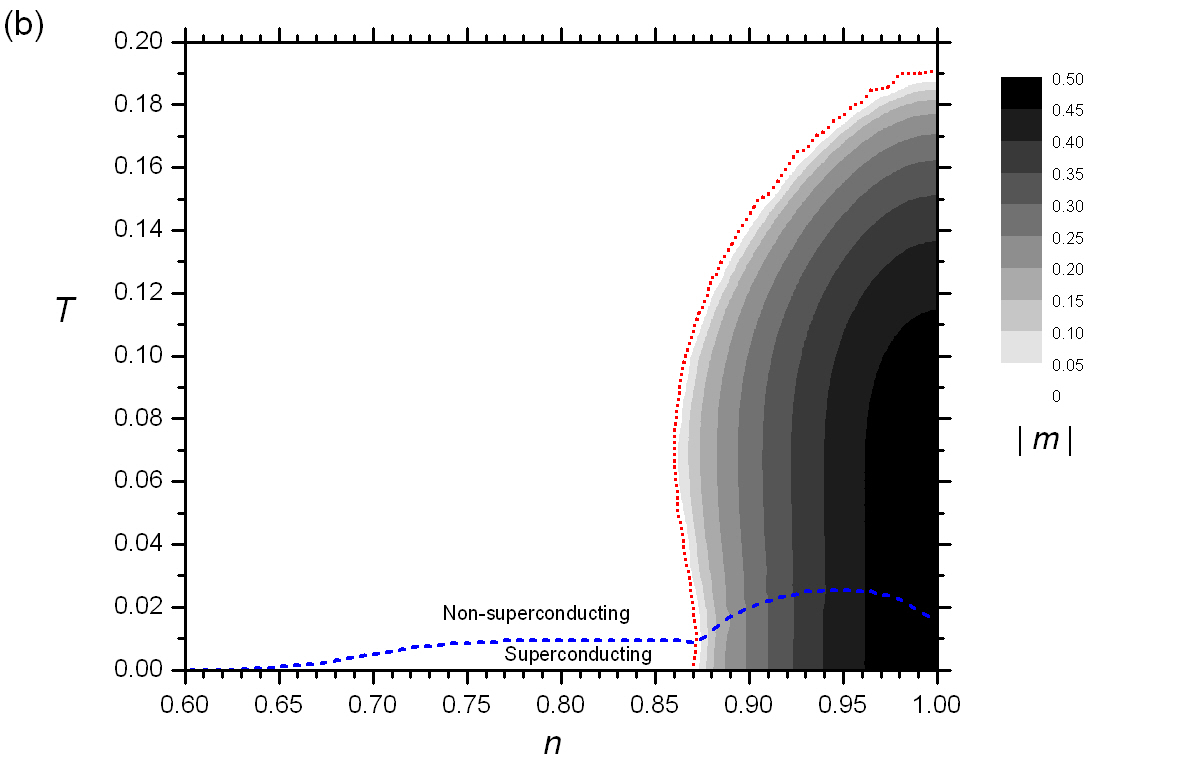} \\

\includegraphics  [scale=0.43]  {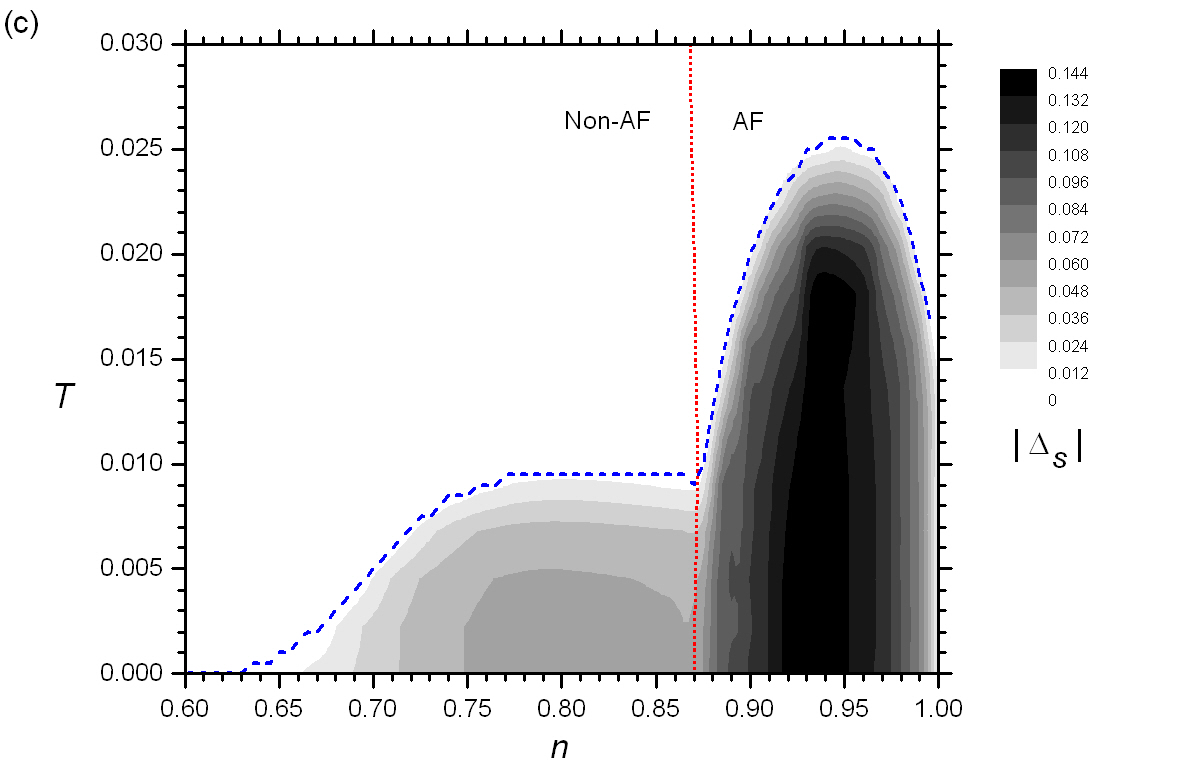} ~~
\includegraphics  [scale=0.42]  {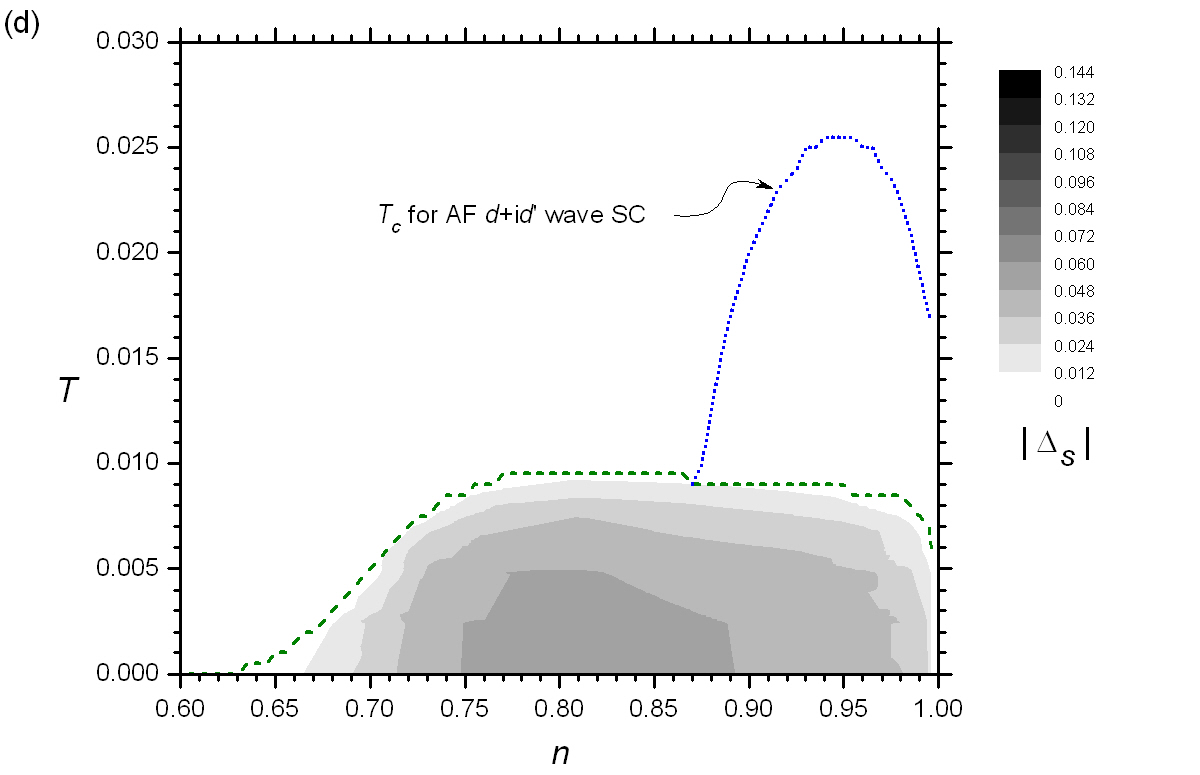} \\

\onehalfspacing

\caption  {
For a system with $(t,t';J) = (-1, 0; 0.5)$, we plot the magnitudes of the AF site spin moment $|m|$ and chiral-$d$ wave superconducting order $|\Delta_s|$ against the band filling fraction $n$ and temperature $T$.
(a) This panel plots the $|m|$ (thick solid line), the $|\Delta_s|$ in an AF state (thin solid line), and the $|\Delta_s|$ in a nonmagnetic state (dotted line) at zero temperature against $n$.
The following panels plot the order parameters on the $n-T$ plane.
The boundaries between AF and nonmagnetic regions, and superconducting and nonsuperconducting regions are also plotted.
(b) This panel plots $|m|$ in the case where there can be a coexisting $|\Delta_s|$.
(c) This panel plots $|\Delta_s|$ in the case where there can be a coexisting $|m|$.
(d) This panel plots the $|\Delta_s|$ in a nonmagnetic system. The $T_c$ for the AF chiral-$d$ wave SC is also included for comparison.
}
\label  {phase-diag}
\end  {figure}

Fig.~\ref {phase-diag}(a) is plotted at zero temperature. The figure plots the coexisting $|m|$ and $|\Delta_s|$ in an AF background, and the $|\Delta_s|$ in a nonmagnetic background against the band filling fraction $n$.
There is a nonvanishing superconducting order $|\Delta_s|$ covering the entire shown band filling range, but it is more noticeable only at $n \gtrsim 0.65$. There is also a prominent AF order $|m|$ in the low doping systems at $n > 0.87$. The most remarkable observation in this figure is in the relative strengths of the $|\Delta_s|$'s in AF and nonmagnetic backgrounds. We see that the $|\Delta_s|$ in an AF state is much stronger. As we will see, this stronger $|\Delta_s|$ is indeed accompanied by a higher superconducting transition temperature in the AF state.

Figs.~\ref {phase-diag}(b) to \ref {phase-diag}(d) plot $|m|$ and $|\Delta_s|$ on the $n-T$ plane with gray scales.
Figs.~\ref {phase-diag}(b) and \ref {phase-diag}(c) plot the $|m|$ and $|\Delta_s|$ respectively in a system where there can be coexisting AFM and SC; whereas Fig.~\ref {phase-diag}(d) plots the $|\Delta_s|$ in a nonmagnetic system.
The boundaries between AF and nonmagnetic regions, superconducting and nonsuperconducting regions are also plotted.
In Fig.~\ref {phase-diag}(b), there is a narrow strip of reentrant zone for the AFM at $0.86 \lesssim n \lesssim 0.87$ and $0.04 \lesssim T \lesssim 0.1$. Besides this reentrant zone, the AF-nonmagnetic boundary is also the locus of the Neel temperature $T_N$.
The AFM can coexist with a SC in the low temperature region at $T \lesssim 0.02$. The superconducting-nonsuperconducting boundary shown in the figure is the locus of the superconducting transition temperature $T_c$.
We can barely see a sign of the presence of SC in the $|m|$ at $T < T_c$.
In Fig.~\ref {phase-diag}(c), there is clearly a disturbance in the $|\Delta_s|$ in the AF regime.
In Fig.~\ref {phase-diag}(d), the $|\Delta_s|$ in a nonmagnetic system is plotted.
Comparing the $|\Delta_s|$'s in AF and nonmagnetic systems, we see a pronounced enhancement in both the magnitude of $|\Delta_s|$ and $T_c$ by the AFM.
The $T_c$ for nonmagnetic states at low doping densities $n \gtrsim 0.9$ is below 0.01, but in an AF background, $T_c$ can be raised to a value as high as 0.025. Regardless of a magnetic or nonmagnetic background, the SC is suppressed at $n$ very close to HBF, since the FS shrinks to a Fermi point and this removes the habitat of the SC.
The Neel temperature $T_N$ and the superconducting transition temperatures $T_c$'s (Figs.~\ref {phase-diag}(b), \ref {phase-diag}(c), and \ref {phase-diag}(d)) have profiles resemble those of the zero temperature $|m|$ and $|\Delta_s|$'s (Fig.~\ref {phase-diag}(a)).

We will see a similar enhancement in the extended-$s$ wave SC by an AFM in Sec.~\ref {xs}. This suggests that the mechanism of the enhancement is based on the band structure, but not the symmetry of the SC. The opening of an AF gap expels the states from the gap and piles up those states on the two sides of the gap, which results in an uprise in the DOS at the Fermi level of a low doping density system. The SC is enhanced by this increase in the DOS.
Noteworthy, though an AFM in a square lattice also leads to a similar enhancement in the DOS, the AFM is found to suppress the $d_{x^2-y^2}$ wave SC in the system \cite {Voo11}.
This is due to the fact that the AF order also renders the one-sublattice structure of the square lattice into a two-sublattice structure. This destroys the FS at the antinodes of a $d_{x^2-y^2}$ wave, and this is a more detrimental factor to the SC. On the other hand, a honeycomb lattice is intrinsically two-sublatticed, and it accommodates an AFM with no destruction in the FS. The SC is simply enhanced by taking the advantage of an increase in the DOS.

\section  {Single particle excitation in antiferromagnetic chiral-$d$ wave states}
\label  {spx}

This section discusses the single particle excitation in the AF chiral-$d$ wave superconducting systems. The excitation is related to the experimental measurements such as the angle-resolved photoemission spectroscopy, specific heat capacity, magnetic penetration depth, etc. We begin with a discussion on the DOS, and then we proceed to discuss the single particle dispersion relation and single particle spectral weight function.

% nodal gaps at two band fillings

\begin {figure}

\includegraphics  [scale=0.45]  {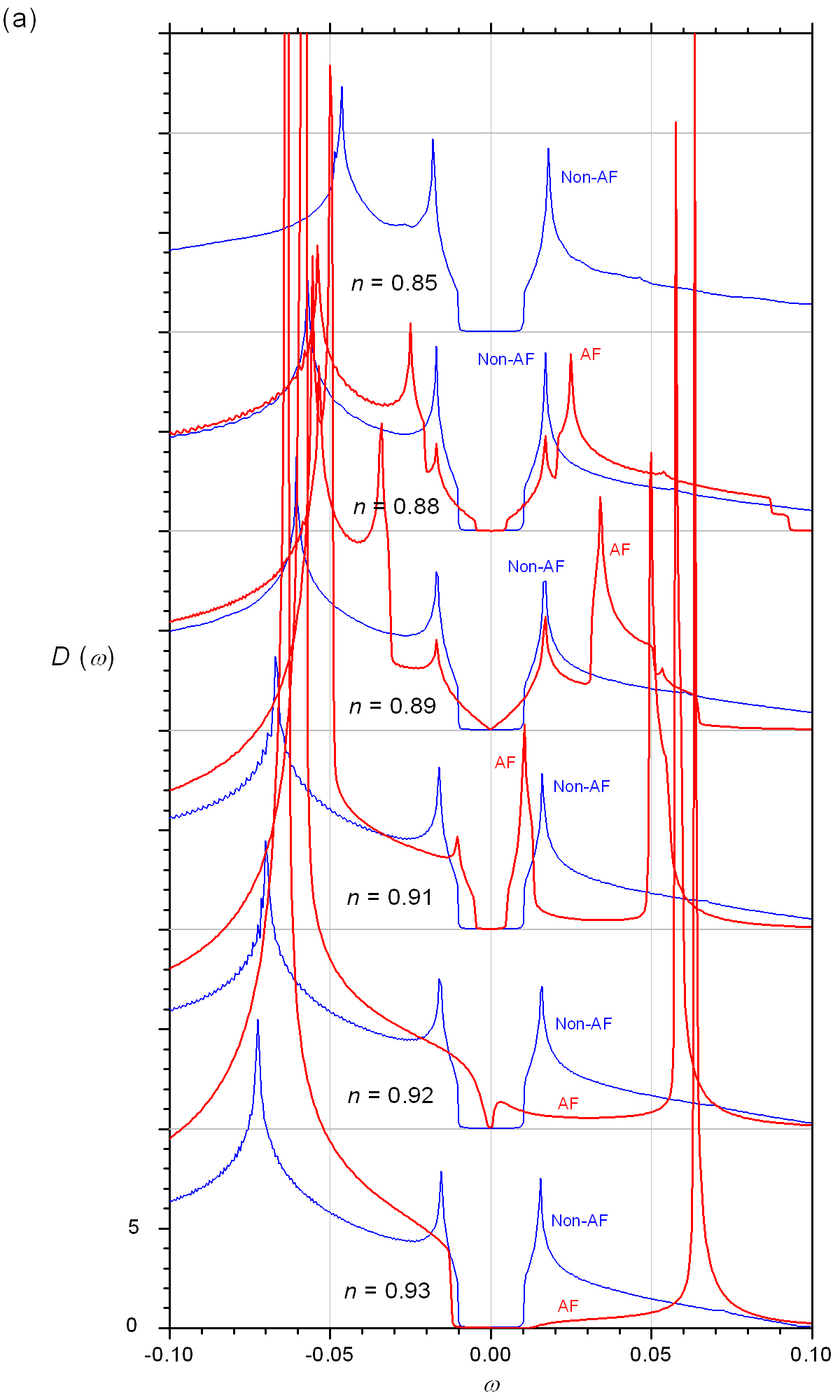}
\includegraphics  [scale=0.45]  {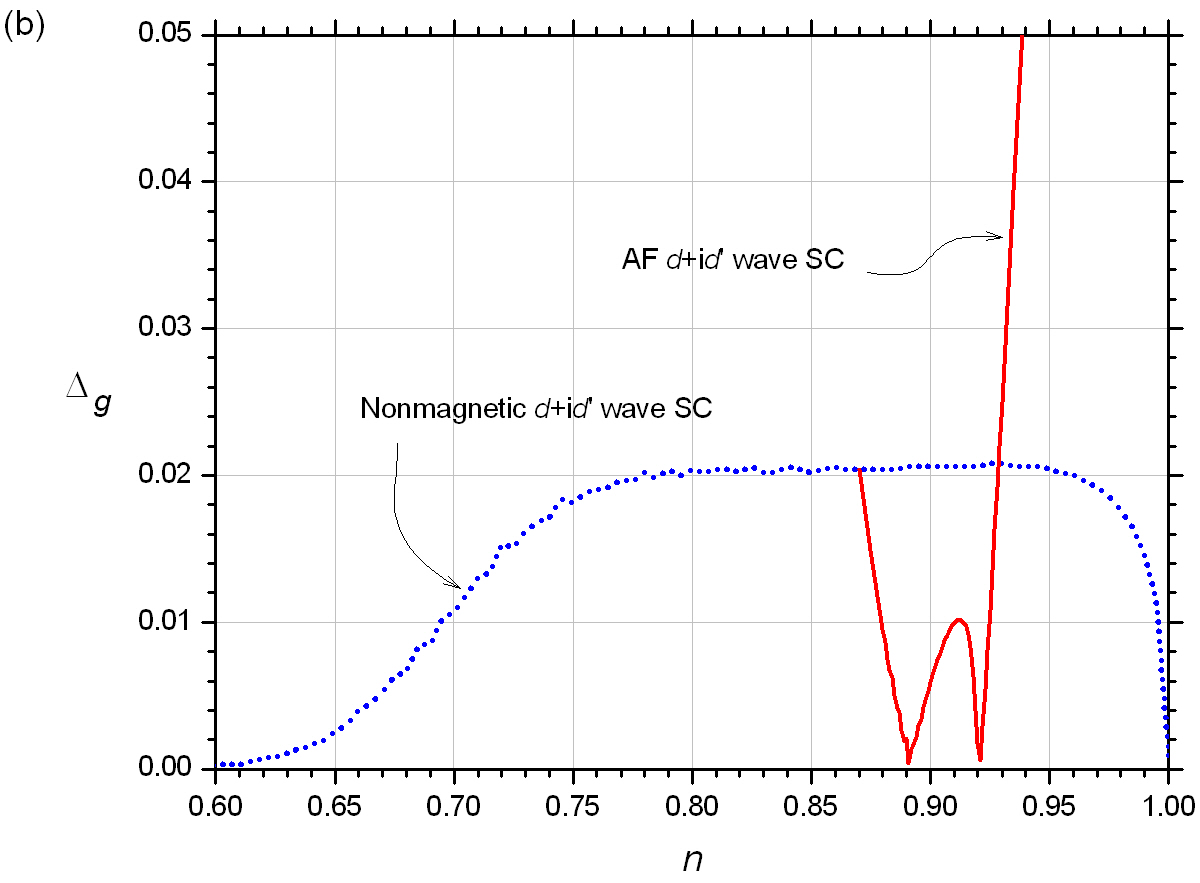} \\

\onehalfspacing

\caption  {
For a system with $(t,t';J) = (-1,0;0.5)$ at zero temperature, we plot the DOS $D (\omega)$ and full energy gap $\Delta_g$ for AF chiral-$d$ wave and nonmagnetic chiral-$d$ wave superconducting states.
(a) This panel plots $D (\omega)$ against the frequency $\omega$, at band filling fractions $n = 0.85$, 0.88, 0.89, 0.91, 0.92, and 0.93. The DOSs are vertically shifted for clarity.
(b) This panel plots $\Delta_g$ against the band filling fraction $n$.
}

\label  {dos-af-did}
\end  {figure}

Fig.~\ref {dos-af-did}(a) plots the DOS in a low frequency window for a few illustrative cases. We consider a system with $(t,t';J) = (1,0;0.5)$ at zero temperature, whose order parameters have been shown in Fig.~\ref {phase-diag}(a).
We plot the DOSs for $n = 0.85$, 0.88, 0.89, 0.91, 0.92, and 0.93.
The system can possess an AFM at band filling fractions $n > 0.87$. At $n > 0.87$, we plot the DOSs for both the AF superconducting and nonmagnetic superconducting states.
Since a chiral-$d$ wave superconducting order is nodeless besides $K$ and $K'$, a doped system is expected to show a full gap at the FL. As expected, we see full gaps in all the nonmagnetic chiral-$d$ wave superconducting states.
But in an AF state, we see that the superconducting gap can become nodal unexpectedly at certain band filling fractions.
We see nodal gaps at $n = 0.89$ and $n = 0.92$. The two nodal gaps have quite different profiles. Their origins differ as well, as we will see later in this section.

% afm splitted coherence peaks :  locations

The presence of coherence peaks in the DOS is an unique feature of superconducting states.
In Fig.~\ref {dos-af-did}(a), we always see a pair of tidy coherence peaks in the DOSs of nonmagnetic superconducting states.
The coherence peaks are due to the expulsion of energy states from the FL by the superconducting potential. In a nonmagnetic superconducting state, those expelled states are simply stacked up at just above the range of this potential, forming the coherence peaks, and thus the locations of the coherence peaks have an evolution trend closely follow the superconducting potential $J |\Delta_s|$.
For example, the coherence peaks in the nonmagnetic DOSs slightly shift to lower frequency when $n$ increases from 0.85 to 0.93 (Fig.~\ref {dos-af-did}(a)), which agrees with the slight decrease in $|\Delta_s|$ in this band filling range (Fig.~\ref {phase-diag}(a)).
When an AFM sets in, it can disturb the DOS quite heavily, and the coherence peaks in the DOSs in the band filling range $0.87 < n < 0.92$ are split up into low frequency and high frequency coherence peaks (Fig.~\ref {dos-af-did}(a)).
The low frequency coherence peaks slowly shift toward zero frequency, while the high frequency coherence peaks shift more rapidly toward higher frequency at approaching HBF.
Nonetheless, the middle point of the low frequency and high frequency coherence peaks  still roughly follows the evolution trend of $J |\Delta_s|$ in the change of $n$.

% afm splitted coherence peaks :   intensities

The pair of low frequency coherence peaks are relatively weaker. Especially when the band filling fraction $n$ is close to 0.92, where those peaks are weakened, rounded off, merged, and disappeared.
At the mergence of the diminishing low frequency coherence peaks, a nodal gap is seen at the FL. Beyond this band filling fraction, the low frequency coherence peaks are not seen.
On the other hand, the pair of high frequency coherence peaks are always present, with growing intensities at approaching HBF.
In the DOSs for $n = 0.91$, 0.92, and 0.93, we see very conspicuous high frequency coherence peaks on the positive frequency side.
The high frequency coherence peak on the negative frequency side are obscured by the overlapping with the van Hove singularity.

% full gap D_g

Fig.~\ref {dos-af-did}(b) plots the full gap $\Delta_g$ against the band filling fraction $n$.
The corresponding system parameters, order parameters, and DOSs are referred to Figs.~\ref {phase-diag}(a) and \ref {dos-af-did}(a).
The full gap $\Delta_g$ is defined as the width of the frequency window in which $D(\omega) = 0$.
The full gap $\Delta_g$ shares the same trend as the pairing amplitude $|\Delta_s|$ only in the nonmagnetic states. In the AF states, $|\Delta_s|$ is always enhanced by the AFM (Fig.~\ref {phase-diag}(a)), but $\Delta_g$ can rise and fall sharply, and even approach zero at $n = 0.89$ and $n = 0.92$. Thus  $\Delta_g$ does not exhibit an intuitive connection with $| \Delta_s |$. Below we proceed to study the single particle dispersion $\varepsilon _{{\bf k} \sigma}$ in order to understand the behavior of $\Delta_g$.

%  plotting parameters in dispersion (fig. 4).

\begin {figure}
\includegraphics  [scale=0.25]  {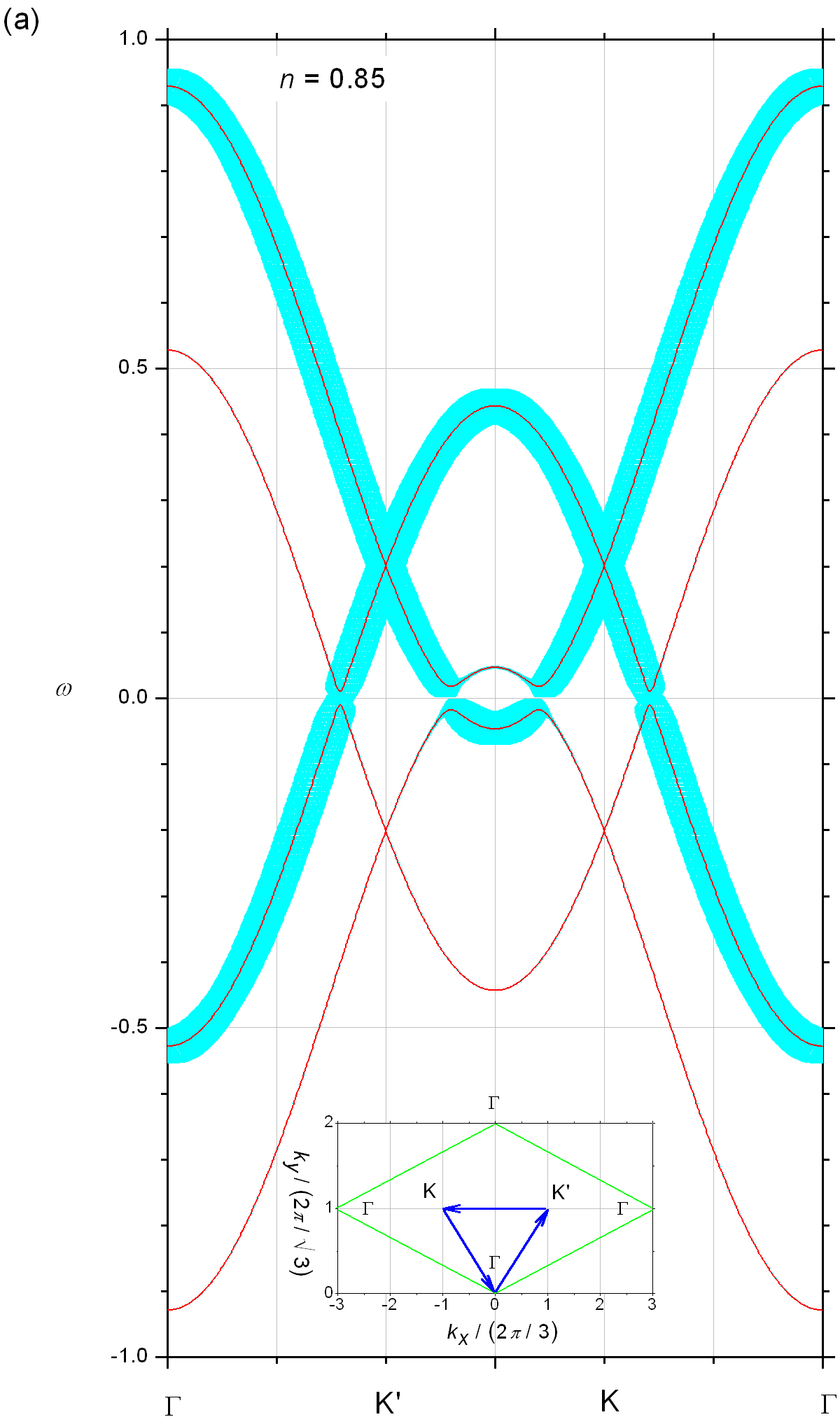}
\includegraphics  [scale=0.25]  {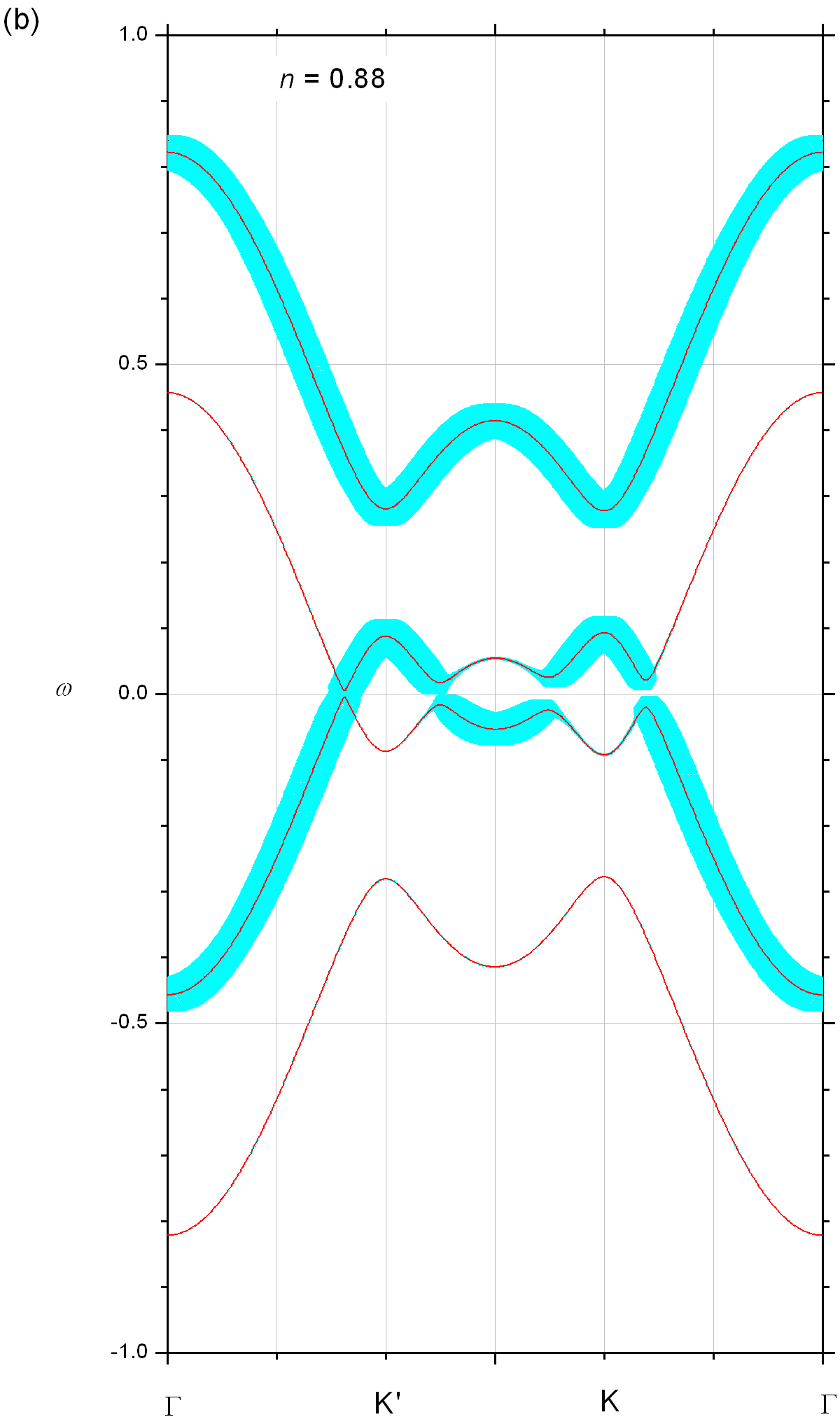}
\includegraphics  [scale=0.25]  {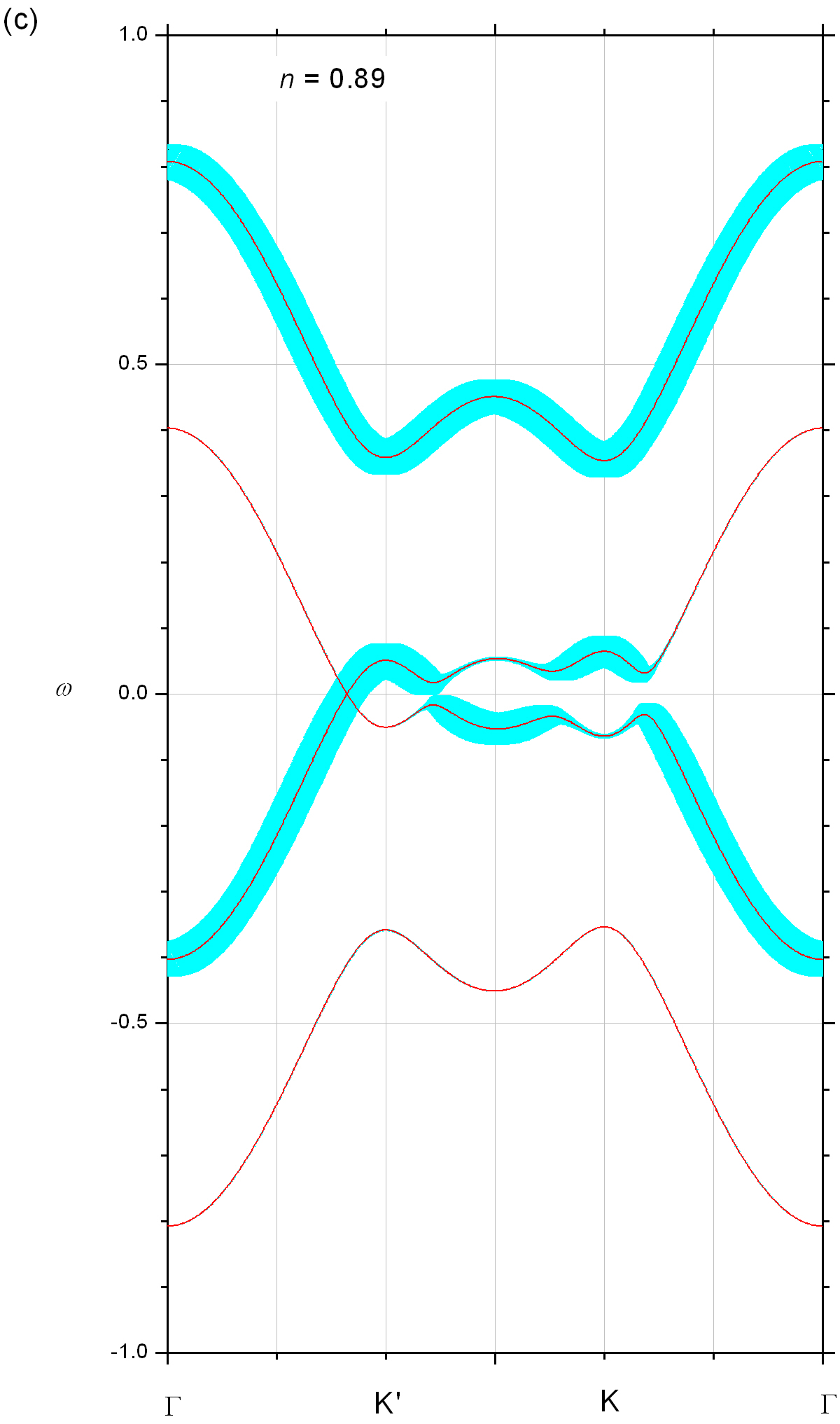} \\

\includegraphics  [scale=0.25]  {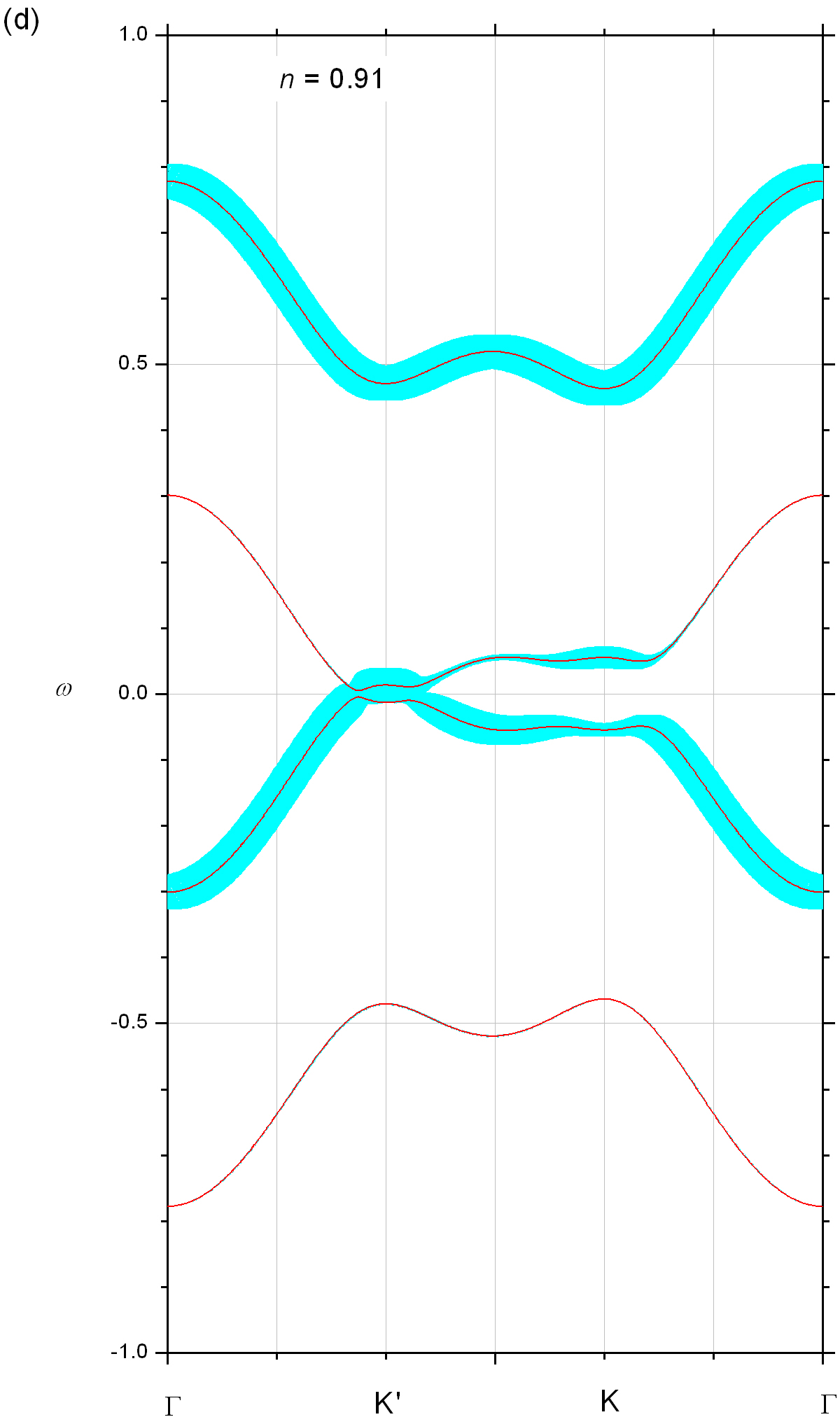}
\includegraphics  [scale=0.25]  {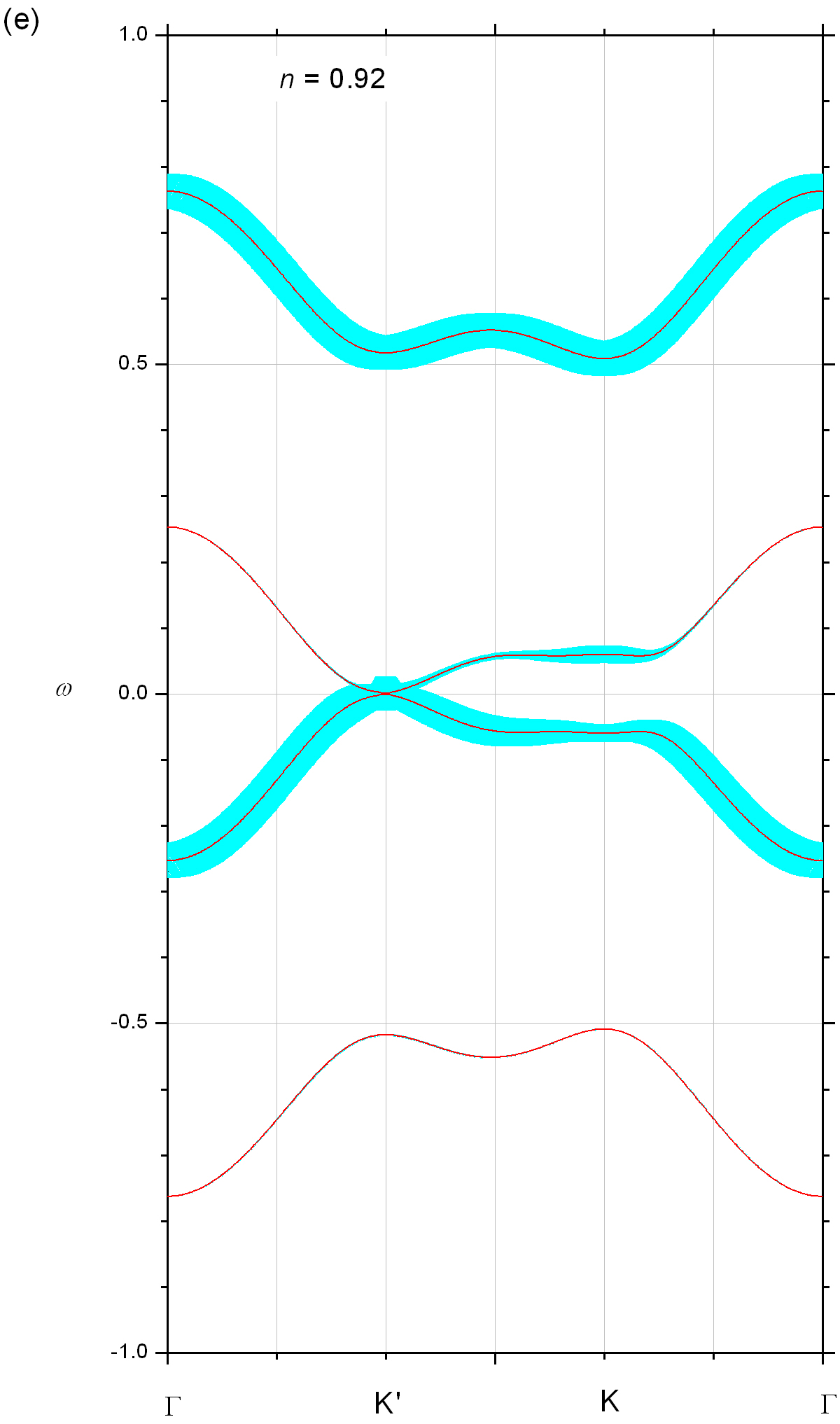}
\includegraphics  [scale=0.25]  {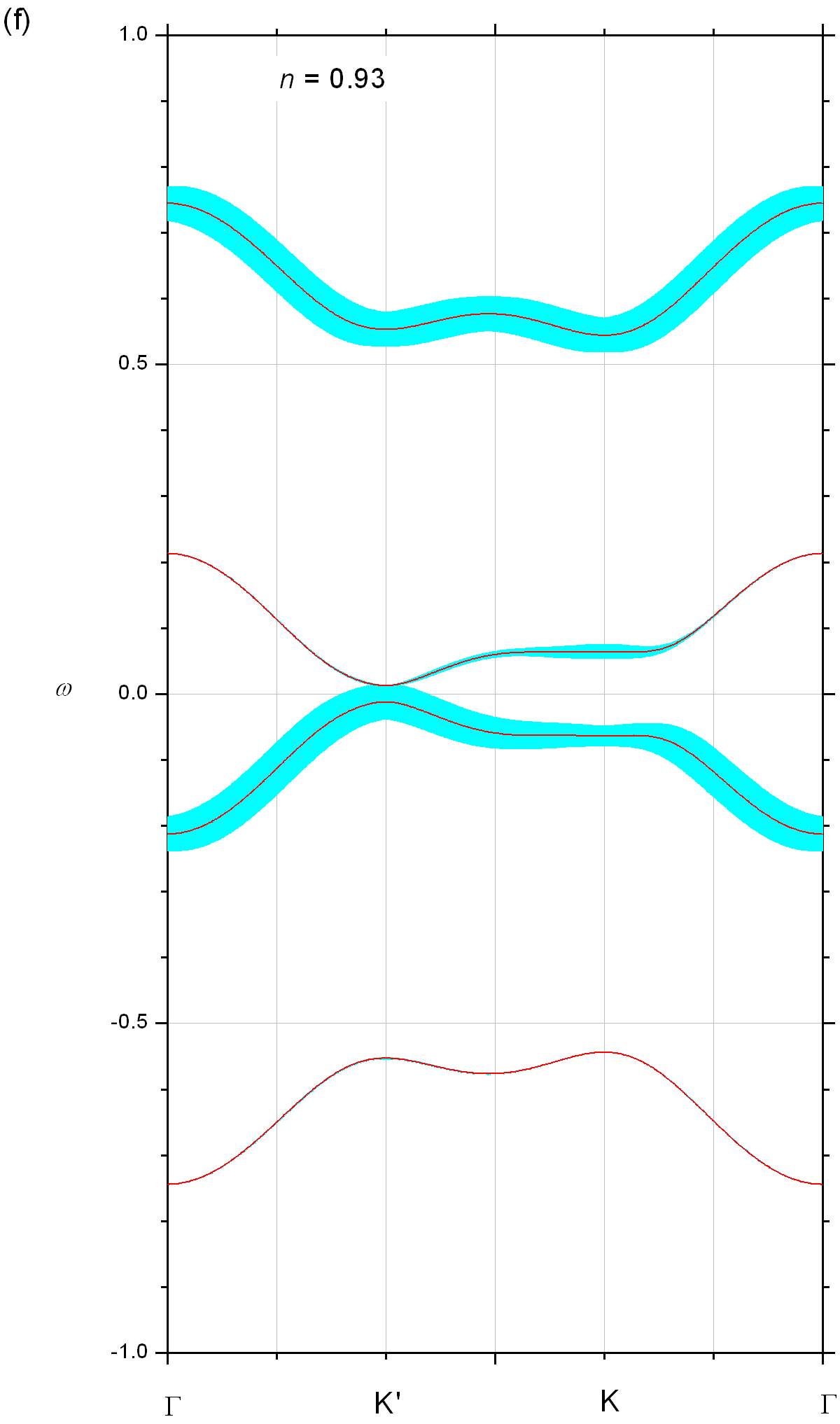} \\

\onehalfspacing

\caption  {
For a zero temperature system with $(t,t';J) = ( 1, 0; 0.5 )$, spin moment $\langle \hat S ^{z, {\rm SB}} _{{\bf i}A} \rangle \geqslant 0$, and superconducting order chirality $\varphi_{\rm sc} = 2 \pi / 3$, we plot the single particle dispersion relation $\varepsilon _{{\bf k, \uparrow}}$ and single particle spectral weight $A ({\bf k},\uparrow; \omega)$ along $\Gamma - {\rm K}'- {\rm K} - \Gamma$. The inset in panel (a) shows the plotting route (the triangle) in a unit BZ (the rhombus).
A dispersion relation branch $\varepsilon _{{\bf k}, \uparrow, j}$ is plotted with a solid line; and its branching weight $a _{{\bf k}, \sigma, j}$  is represented by a proportionally drawn local thickness of the line envelop on the dispersion line, where $j = 1$, 2, 3, and 4.  We show the results for (a) $n = 0.85$, (b) $n = 0.88$, (c) $n = 0.89$, (d) $n = 0.91$, (e) $n = 0.92$, and (f) $n = 0.93$.
}

\label  {aksw}
\end  {figure}

Fig.~\ref {aksw} plots the single particle dispersion relation $\varepsilon _{{\bf k, \uparrow}}$ along the route $\Gamma - K' - K - \Gamma$ (this route is shown in the inset of Fig.~\ref {aksw}(a)).
The systems have $(t,t';J) = ( 1, 0; 0.5 )$, spin moment $\langle \hat S ^{z, {\rm SB}} _{{\bf i}A} \rangle \geqslant 0$, chiral-$d$ wave superconducting order with chirality $\varphi_{\rm sc} = 2 \pi / 3$, and the systems are at zero temperature.
We consider a spin-$\uparrow$ electron.
There are four weighted dispersion branches, each dispersion branch $\varepsilon _{{\bf k}, \uparrow, j}$ carries a weight $a _{{\bf k}, \uparrow, j}$, where $j = 1$, 2, 3, and 4.
When the weight of a dispersion branch vanishes, the branch is physically absent.
Figs.~\ref {aksw}(a) to \ref {aksw}(f) are plotted at the band filling fractions whose DOSs have been presented in Fig.~\ref {dos-af-did}(a).

%  overall features :   Mott shrunk band width, two-branches, AF gap, parabolic dips-peaks.

Fig.~\ref {aksw} covers the single particle dispersion relations $\varepsilon _{{\bf k, \uparrow}}$ in both the nonmagnetic and AF superconducting states.
In a SB MFT, the Mott nature of a system is modeled by a reduction of the band width by a factor of $\delta$, that results in the shrinkage in the overall band widths  from Fig.~\ref {aksw}(a) to \ref {aksw}(f).
The dispersions are developed from a two-branches structure.
The free band dispersion of a honeycomb lattice consists of two branches due to its two-sublattices structure.
An AFM does not alter the two-branches structure of the dispersion, but nonetheless modifies it by opening an AF gap at $K$ and $K'$, thereby rounding the upward and downward touching Dirac cones into detached parabolic peaks and dips.
The AF gap has a width about the order of $3 J |m|$, where the factor 3 is due to the coordination number of a honeycomb lattice site.
At $n = 0.93$, we have $|m| \simeq 0.37$, which results in an AF gap with a width about 0.5 (Fig.~\ref {aksw}(f)).
A SC doubles the number of dispersion branches into four, but the two additional branches are physically significant only when they are within a frequency scale of the superconducting potential $J |\Delta_s|$ about the FL, or $|\omega| < J |\Delta_s|$.
Referring to Fig.~\ref {phase-diag}(a), the superconducting potential $J |\Delta_s|$ is not more than 0.1 in those dispersions in Fig.~\ref {aksw}.
An inspection on the distribution of the spectral weights on the dispersion branches in the figure, we see that in most part of the route the weights are equally distributed on only two of the branches, or the two-branches structure of the dispersion relation is maintained.

%  overall features :   SC fragmentation to 3-branches, asymmetric K '/ K

\begin {figure}

\includegraphics  [scale=0.4]  {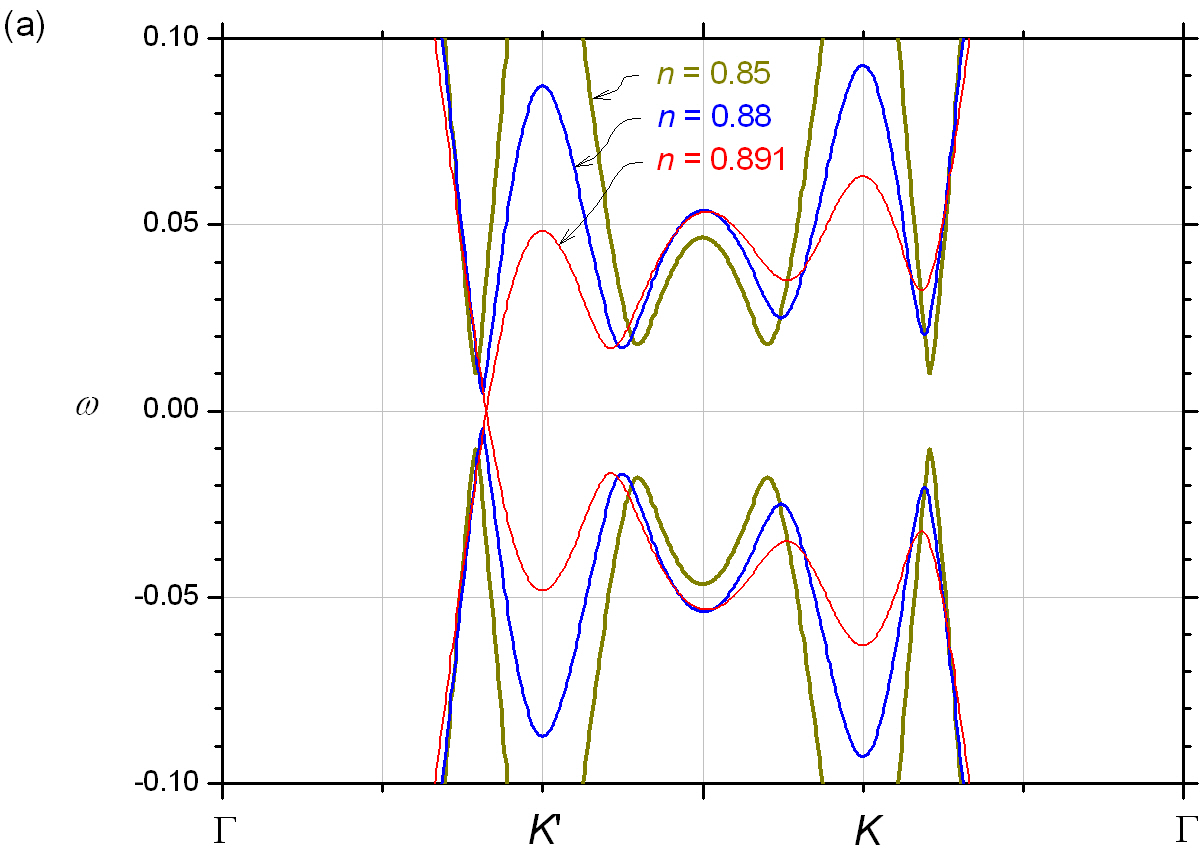} ~~~~~~
\includegraphics  [scale=0.4]  {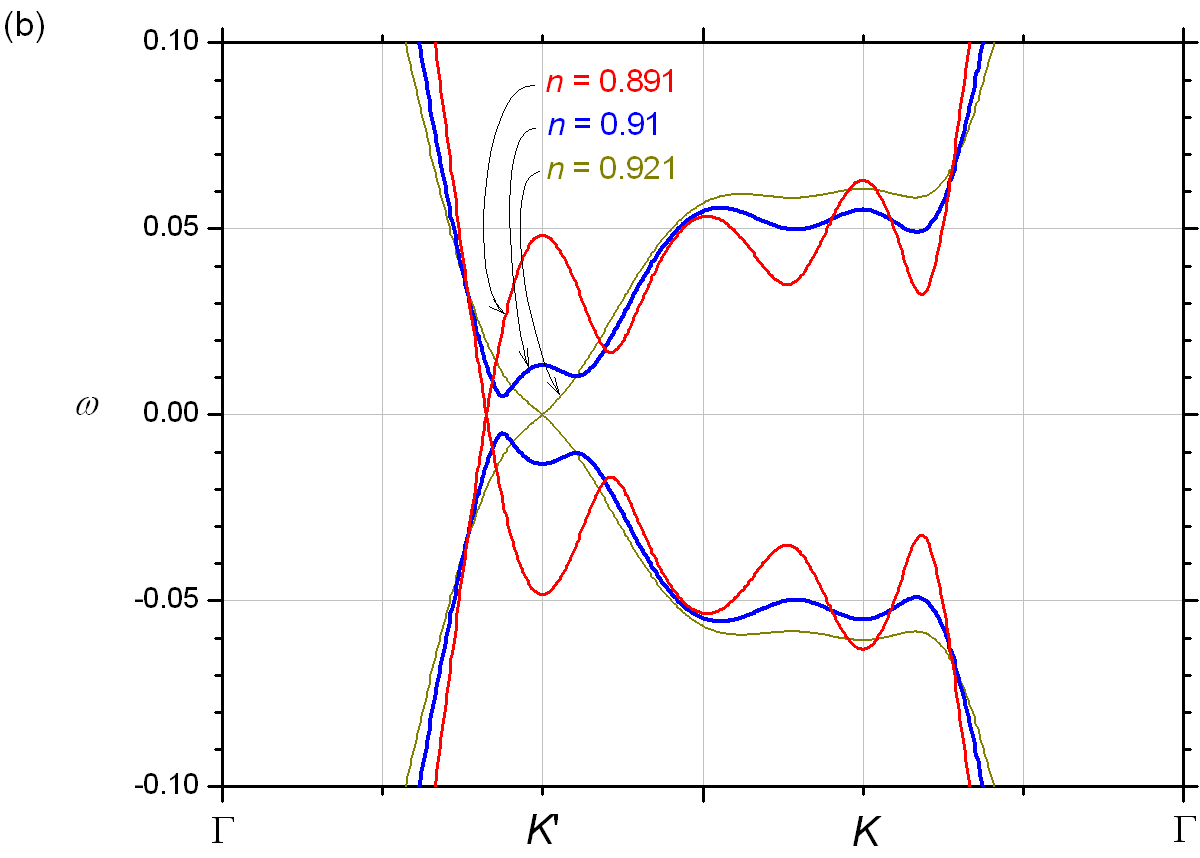} \\

\includegraphics  [scale=0.4]  {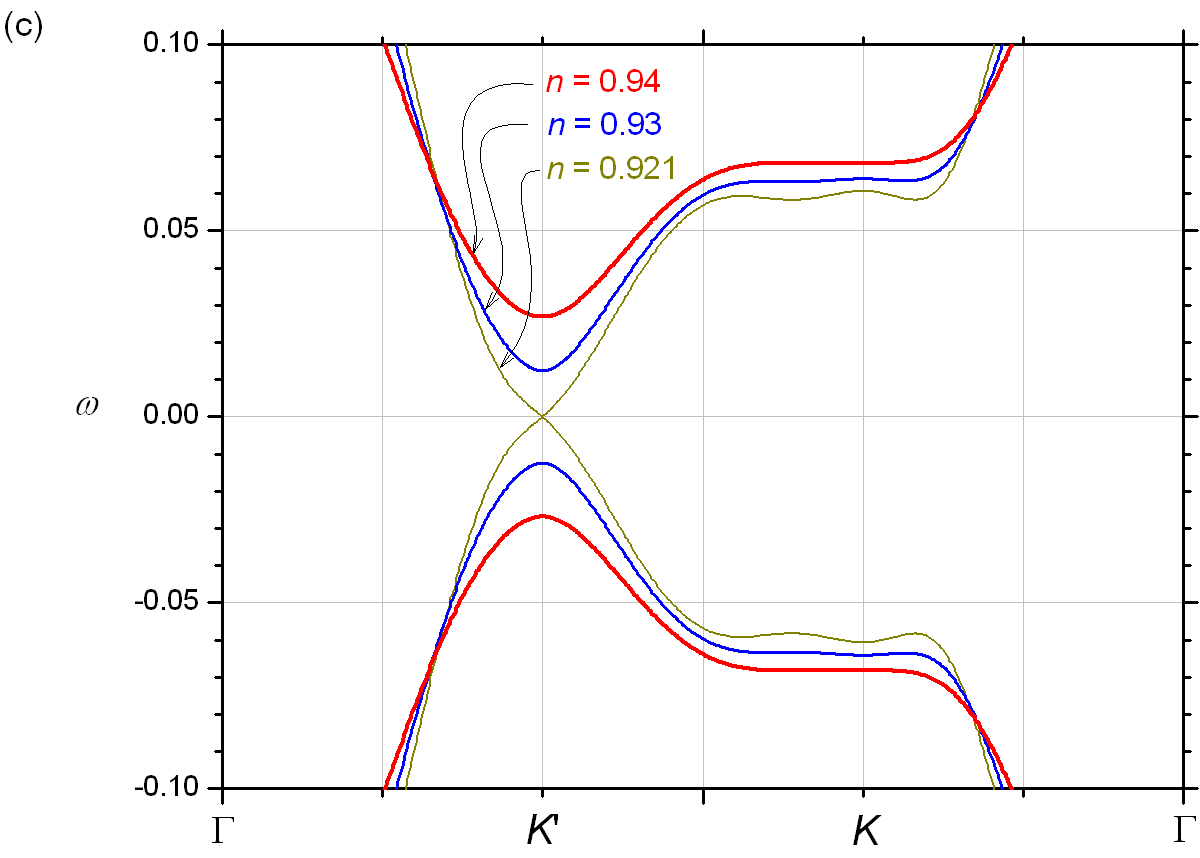} ~~~~~~
\includegraphics  [scale=0.4]  {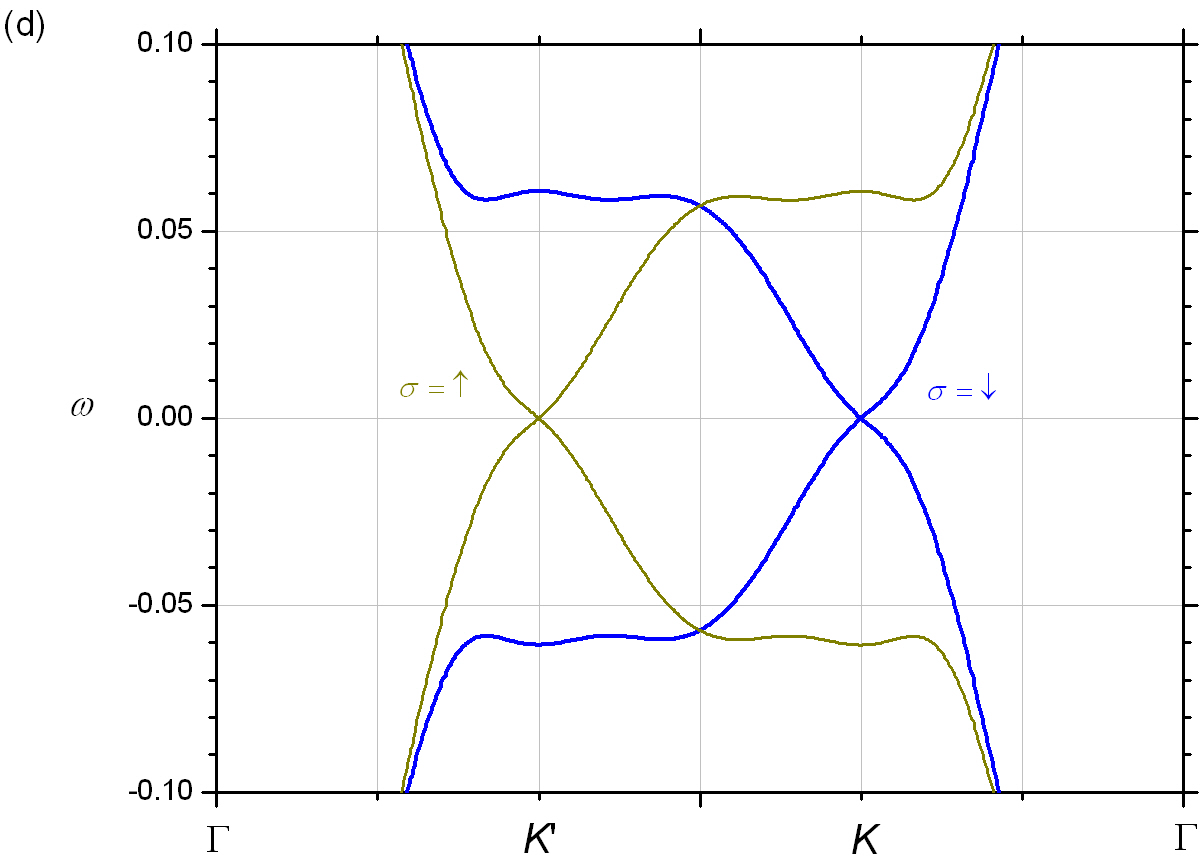} \\

\includegraphics  [scale=0.65]  {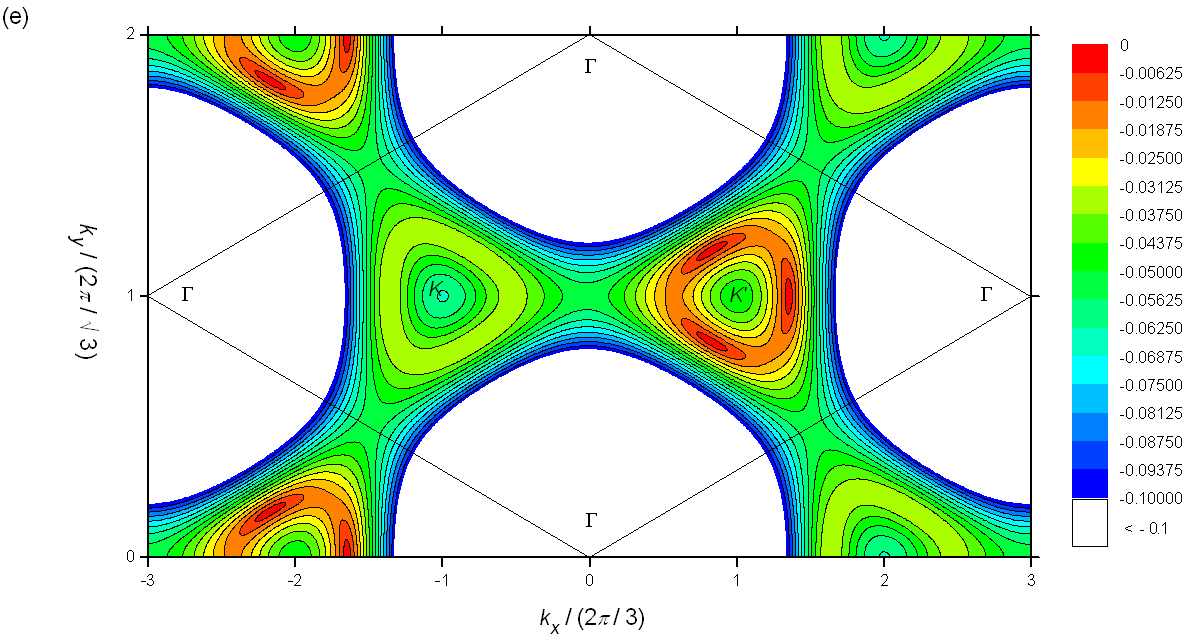} \\

\onehalfspacing

\caption  {
The system parameters are referred to Fig.~\ref {aksw}.
Along $\Gamma - K'- K - \Gamma$, the low frequency part of $\varepsilon _{{\bf k, \uparrow}}$ is plotted for (a) $n=0.85$, 0.88, and 0.891; (b) $n = 0.891$, 0.91, and 0.921;  and (c) $n=0.921$, 0.93, and 0.94. (d) The low frequency part of $\varepsilon _{{\bf k, \uparrow}}$ and $\varepsilon _{{\bf k, \downarrow}}$ for $n = 0.921$ are juxtaposed. (e) For $n = 0.891$, the $\varepsilon _{{\bf k, \uparrow}}$ in the frequency range $- 0.1 < \omega < 0$ is plotted on the BZ. The rhombus frames a unit BZ.
}

\label  {aksw_zoom}
\end  {figure}

A dispersion has effectively more than two branches only
when some of the dispersion branches enter the frequency window $|\omega| < J |\Delta_s|$, or $|\omega| \lesssim 0.1$ in Fig.~\ref {aksw}.
For examples, in Figs.~\ref {aksw}(a), \ref {aksw}(b),  and \ref {aksw}(c), we see that $\varepsilon _{{\bf k}, \uparrow, 2}$ and $\varepsilon _{{\bf k}, \uparrow, 3}$ can be equally weighted only at their very low frequency sections, found located on the two sides of $K'$ and $K$.
In these sections of the routes, the dispersions have effectively three branches.
Fig.~\ref {aksw}(a) shows the dispersion for a superconducting but nonmagnetic system at $n = 0.85$. We see a superconducting gap with a width about 0.01 at the FL.
The Dirac cones at $K'$ and $K$ high above the FL are intact.
Figs.~\ref {aksw}(b) to \ref {aksw}(f) show the dispersions with AF gaps at $K'$ and $K$.
In a hole doped system, the expulsion of electronic states from an AF gap is also a compression of electronic states toward the FL. This piling up of states at the FL explains the enhancement of the superconducting order $\Delta_s$ by the AFM (Fig.~\ref {phase-diag}(a)).
The parabolic dips at $K'$ and $K$ are always well preserved, since they are well above the FL.
Whereas, the parabolic peaks at $K'$ and $K$ are distorted by the SC, since they are near the FL.
The low frequency dispersion branches are also fragmented into emergent branches by the SC,  when they get close to the FL, via a division of the spectral weight.
Note that, when a system is simultaneously AF and superconducting, there is an asymmetry in $K'$ and $K$ in the low frequency dispersion.
Overall speaking, this asymmetry becomes more pronounced at approaching HBF (e.g., see Figs.~\ref {aksw}(d), \ref {aksw}(e), and \ref {aksw}(f)). The details in the low frequency sector will be studied in Fig.~\ref {aksw_zoom}.

%  afm onset. asymmetry in K and K'. closed gap.

Fig.~\ref {aksw_zoom} plots the low frequency part of the dispersion relation $\varepsilon _{{\bf k, \uparrow}}$.
We will elaborate on the asymmetry effect in the dispersion relation between $K'$ and $K$, in the simultaneous presence of AFM and chiral-$d$ SC.
Fig.~\ref {aksw_zoom}(a) plots the dispersion relations at a few band filling fractions near the onset of the AFM.
We see that the nonmagnetic superconducting dispersion at $n = 0.85$ is symmetric in $K'$ and $K$. In contrast, the AF superconducting dispersion at a slightly increased band filling fraction $n = 0.88$ exhibits an asymmetry between $K'$ and $K$. Referring to the DOS for the $n = 0.88$ system in Fig.~\ref {dos-af-did}(a), we see that this asymmetry is the cause of the splitting of the coherence peaks.
The asymmetry grows as the system approaches HBF. At $n = 0.891$, the dispersion around $K'$ and $K$ can be so astonishing that there is a closed gap near $K'$, but a widened gap near $K$. We will see a closed gap at $K'$ again at $n = 0.921$ in the later discussion (see Fig.~\ref {aksw_zoom}(b)).

%  intermediate afm. two closing gaps.

Fig.~\ref {aksw_zoom}(b) plots the low frequency part of $\varepsilon _{{\bf k, \uparrow}}$ at band filling fractions where the AFM has grown into an intermediate strength.
We see that going beyond $n = 0.891$, the gap is reopened at $n = 0.91$, and then it is closed again at $n= 0.921$. The closing of gap at $n= 0.921$ occurs right on $K'$. This gap closing at $K'$ is also accompanied by a gap widening near $K$.
From the emergence of AFM at $n = 0.88$ to the second gap closing at $n= 0.921$, we note that the extrema of the dispersion surrounding $K'$ actually converge into $K'$ and decrease their energy at the same time, and end up in a coalescence of the extrema and a gap node at $K'$.
The development of these dispersion extrema surrounding $K'$ results in the softening and disappearance of the low frequency coherence peaks we see in Fig.~\ref {dos-af-did}(a).
When a gap closes, the dispersion at the gap node is always in a Dirac form, regardless of the location of the gap closing (see the dispersions for $n=0.891$ and 0.921 in Figs.~\ref {aksw_zoom}(a) and \ref {aksw_zoom}(b) respectively).

% mechanism of gap closing at K and K'

The gap closing at $n = 0.921$ which occurs right on $K'$, can be readily understood by investigating the determinant $D _{{\bf k}, \sigma}$ (see Eq.~\ref {det}), whose vanishing implies a closing gap.
First we note that $\varepsilon ^{AB} _{\bf k} = \Delta^{AB} _{{\bf k} , \sigma } \Delta ^{AB} _{- {\bf k} , { \bar \sigma } } = 0$ at both $K'$ and $K$, which reduces the determinant to $D_{{\bf k},\sigma} =
( {\varepsilon ' _{ {\bf k} A \sigma } }
{\varepsilon ' _{{ \bf k} A \bar \sigma } } ) ^ 2
+ | {\varepsilon ' _{ {\bf k} A \sigma } } \Delta ^{AB} _{- {\bf k} , { \bar \sigma } } | ^ 2
+ | {\varepsilon ' _{ {\bf k} A \bar \sigma } }  \Delta^{AB} _{{\bf k} , \sigma } | ^ 2$
at these two points.
Considering $t' = 0$, we have $\varepsilon ' _{ {\bf k} A \sigma } = - 3 \eta_\sigma J m / 2 - \mu$.
Since we are considering $\eta_\sigma m > 0$ and a hole doped system where $\mu < 0$, we have $\varepsilon ' _{ {\bf k} A \sigma } = - 3 J |m| / 2 + |\mu|$, where a growing $|m|$ can result in $\varepsilon ' _{ {\bf k} A \sigma } = 0$ \cite {Voo22ps}. At this vanishing, $D_{{\bf k},\sigma}$ is further reduced to $D_{{\bf k},\sigma} = | {\varepsilon ' _{ {\bf k} A \bar \sigma } }  \Delta^{AB} _{{\bf k} , \sigma } | ^ 2$.
We consider $\varphi_{\rm sc} = 2 \pi / 3$, which has $\Delta^{AB} _{{\bf k} , \sigma} = 0$ at $K'$, and that results in $D_{{\bf k},\sigma} = 0$ and a closing gap at $K'$.
Note that $\varepsilon ' _{ {\bf k} A \sigma }$ and $\varepsilon ' _{ {\bf k} A \bar \sigma }$ are always nonvanishing in a doped and nonmagnetic system, where $m = 0$ and $|\mu| > 0$.

%  low frequency dispersion  :   strong afm. parabolic peak / dip

Fig.~\ref {aksw_zoom}(c) plots the low frequency part of $\varepsilon _{{\bf k, \uparrow}}$ in the strong AFM regime. We plot the $\varepsilon _{{\bf k, \uparrow}}$ for band filling fractions beyond $n = 0.921$, at which there is a gap closing. The dispersion has a rather simple evolution in this regime.
At $n > 0.921$, the dispersion has a parabolic peak and a parabolic dip pointing each other at $K'$. When the band filling fraction approaches HBF, the gap at $K'$ is monotonically widened, while the gap near $K$ is saturated. The dispersion near $K$ is also progressively flattened, giving rise to strong peaks in the DOS (see Fig.~\ref {dos-af-did}(a)).

% low frequency dispersion  :   spin-inversion and K-K' switch

Fig.~\ref {aksw_zoom}(d) juxtaposes the two dispersion relations $\varepsilon _{{\bf k, \uparrow}}$ and $\varepsilon _{{\bf k, \downarrow}}$ for $n = 0.921$. We see a switch of the locations of the gap nodes between $K'$ and $K$.
The gap closing of $\varepsilon _{{\bf k, \downarrow}}$ at $K$ has a reason parallel to that for the gap closing of $\varepsilon _{{\bf k, \uparrow}}$ at $K'$. We start from the reduced $D_{{\bf k},\sigma} =
( {\varepsilon ' _{ {\bf k} A \sigma } }
{\varepsilon ' _{{ \bf k} A \bar \sigma } } ) ^ 2
+ | {\varepsilon ' _{ {\bf k} A \sigma } } \Delta ^{AB} _{- {\bf k} , { \bar \sigma } } | ^ 2
+ | {\varepsilon ' _{ {\bf k} A \bar \sigma } }  \Delta^{AB} _{{\bf k} , \sigma } | ^ 2$
at $K'$ and $K$.
For a hole doped system with $t' = 0$ and $\eta_\sigma m < 0$, we have $\varepsilon ' _{ {\bf k} A \bar \sigma } = - 3 J |m| / 2 + |\mu|$, where a growing $|m|$ at approaching HBF can lead to $\varepsilon ' _{ {\bf k} A \bar \sigma } = 0$, and results in $D_{{\bf k},\sigma} =  | {\varepsilon ' _{ {\bf k} A \sigma } } \Delta ^{AB} _{- {\bf k} , { \bar \sigma } } | ^ 2$ at $K'$ and $K$.
For the chirality $\varphi_{\rm sc} = 2 \pi / 3$, the superconducting potential node $\Delta ^{AB} _{- {\bf k} , { \bar \sigma } }  = 0$ at $K$ results in $D_{{\bf k},\sigma} = 0$ and a closing gap at $K$.
Otherwise, we may also understand the switch of the gap nodes in Fig.~\ref {aksw_zoom}(d) as the result of a switch in the sign of $\varphi_{\rm sc} \eta_\sigma m$ (see Sec.~\ref {form}).
We have considered systems with $\varphi_{\rm sc} > 0$ and $m > 0$, where $\varphi_{\rm sc} \eta_\uparrow m > 0$ and $\varphi_{\rm sc} \eta_\downarrow m < 0$.
Therefore, we have $\varepsilon _{{\bf k, \uparrow}} = \varepsilon _{\bf k} ^ +$ and $\varepsilon _{{\bf k, \downarrow}} = \varepsilon _{\bf k} ^ -$.
An interchange of the wave vector axes $( k_u, k_v ) \rightarrow (k_v, k_u)$ transforms
one dispersion to the other, i.e., $\varepsilon _{{\bf k, \uparrow}} | _ { ( k_u, k_v ) \rightarrow (k_v, k_u) }= \varepsilon _{{\bf k, \downarrow}}$, hence switches the gap node between $K'$ and $K$.

% a mean to obtain spin-polarized electrons

The asymmetric dispersions $\varepsilon _{{\bf k, \uparrow}}$ and $\varepsilon _{{\bf k, \downarrow}}$ in an AF superconducting state might suggest a mean to obtain spin-polarized electrons.  In circumstances such as that in Fig.~\ref {aksw_zoom}(d), the up and down spin electrons can be quite well separated in frequency and momentum.
Therefore we may readily obtain spin polarized electrons by, e.g.,  selectively ejecting electrons by photoemission effect from the  neighborhoods of $K'$ and $K$. Though the asymmetry in the dispersion in $K'$ and $K$ is present throughout the AF regime, the frequency difference $| \varepsilon _{{\bf k, \uparrow}} - \varepsilon _{{\bf k, \downarrow}} |$ is maximized in the intermediate AFM regime near $n \simeq 0.92$.

%  low frequency dispersion  :   triplet nodes

In closing this section, we prepare a few remarks for the dispersion nodes. We see that the nodeless chiral-$d$ wave superconducting order can have closing dispersion gaps at two band filling fractions in the AF regime.
For the gap closing at $n = 0.921$, the gap closes right on $K'$ or $K$, and the low frequency dispersion is a Dirac cone roughly isotropic in the $K'-\Gamma$ and $K'-K$ directions (Fig.~\ref {aksw_zoom}(d)).
For the gap closing at $n = 0.891$, we plot the $\varepsilon _{{\bf k, \uparrow}}$ in the frequency range $- 0.1 < \omega < 0$ on a two-dimensional BZ (Fig.~\ref {aksw_zoom}(e)). The dispersion nodes are seen to distribute in the form of a triplet encircling $K'$. The offset of the gap nodes from $K'$ and $K$ hinders a more detailed analytic analysis of the gap closing. The low frequency dispersions at the nodes are Dirac cones with very anisotropic velocities.

\section  {Comparison with antiferromagnetic extended-$s$ wave states}
\label  {xs}

\begin {figure}

\includegraphics  [scale=0.4]  {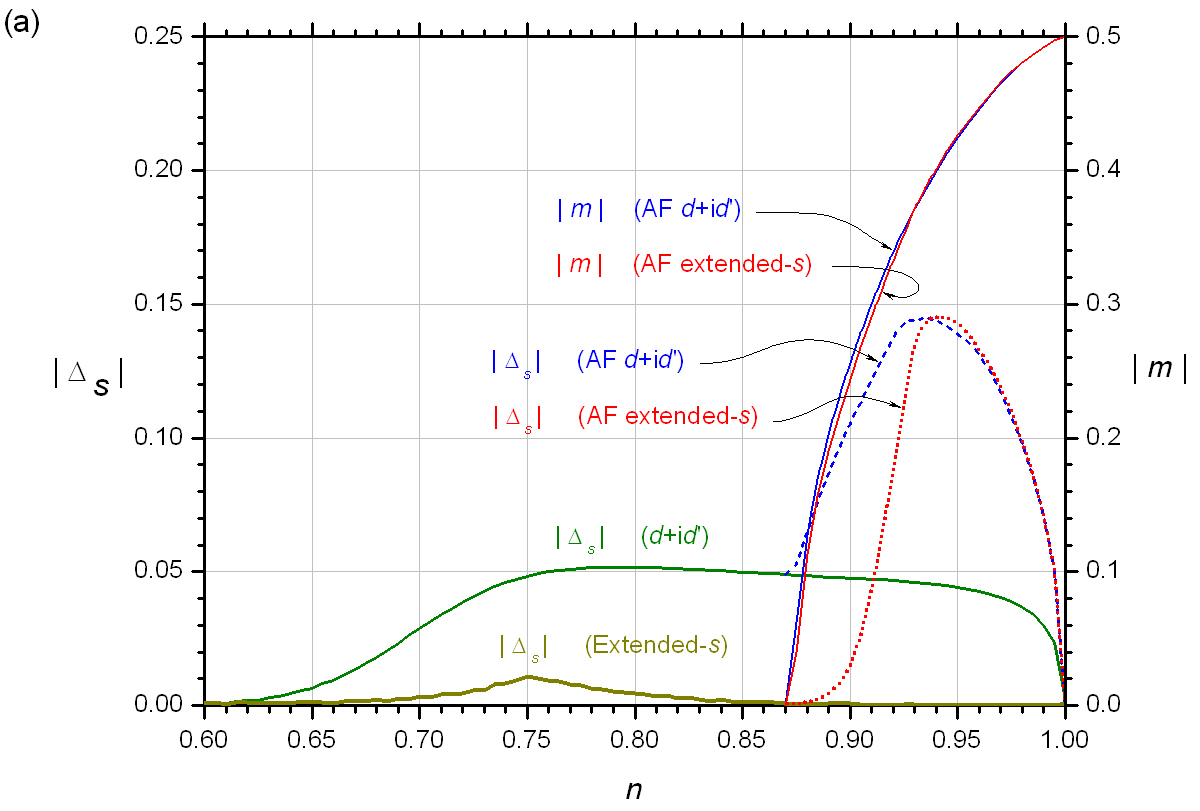} ~~~~~~
\includegraphics  [scale=0.36]  {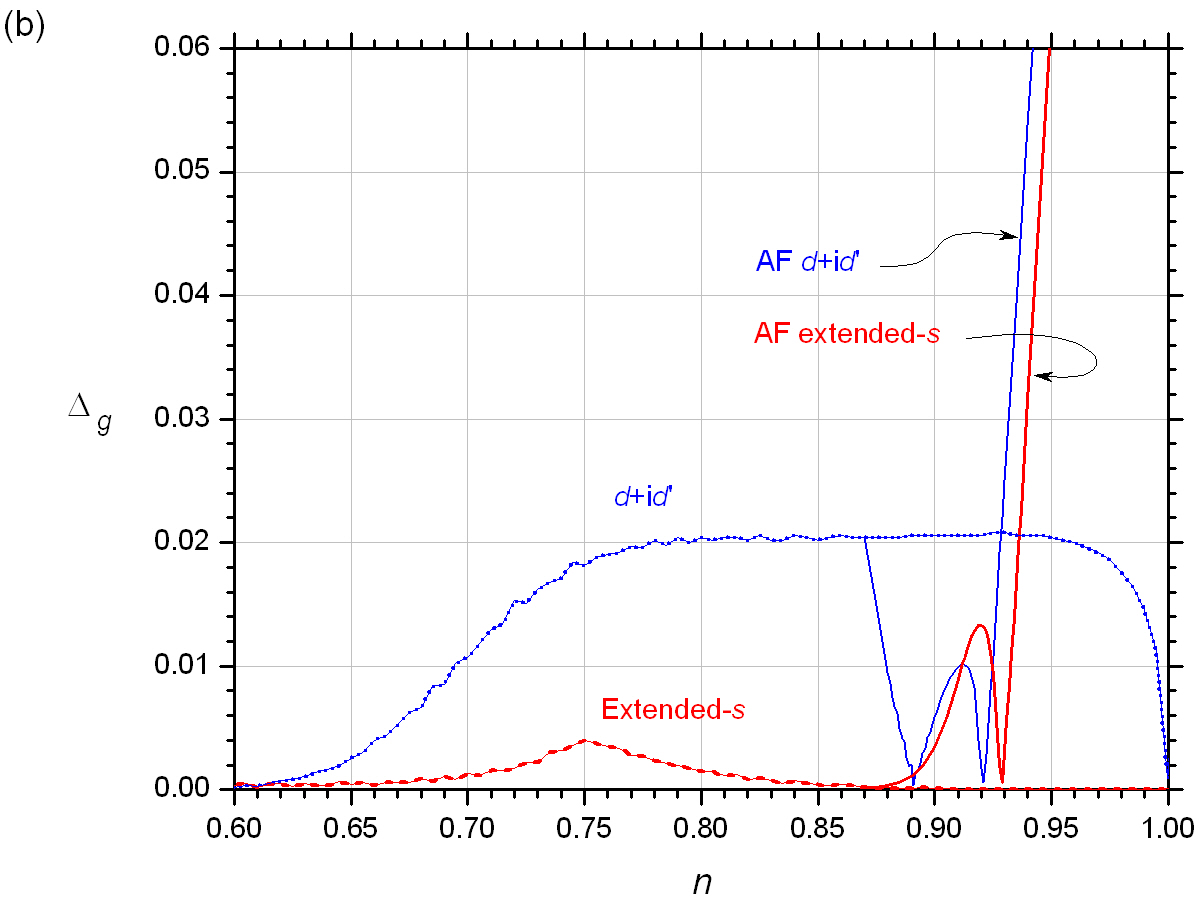} \\

\includegraphics  [scale=0.36]  {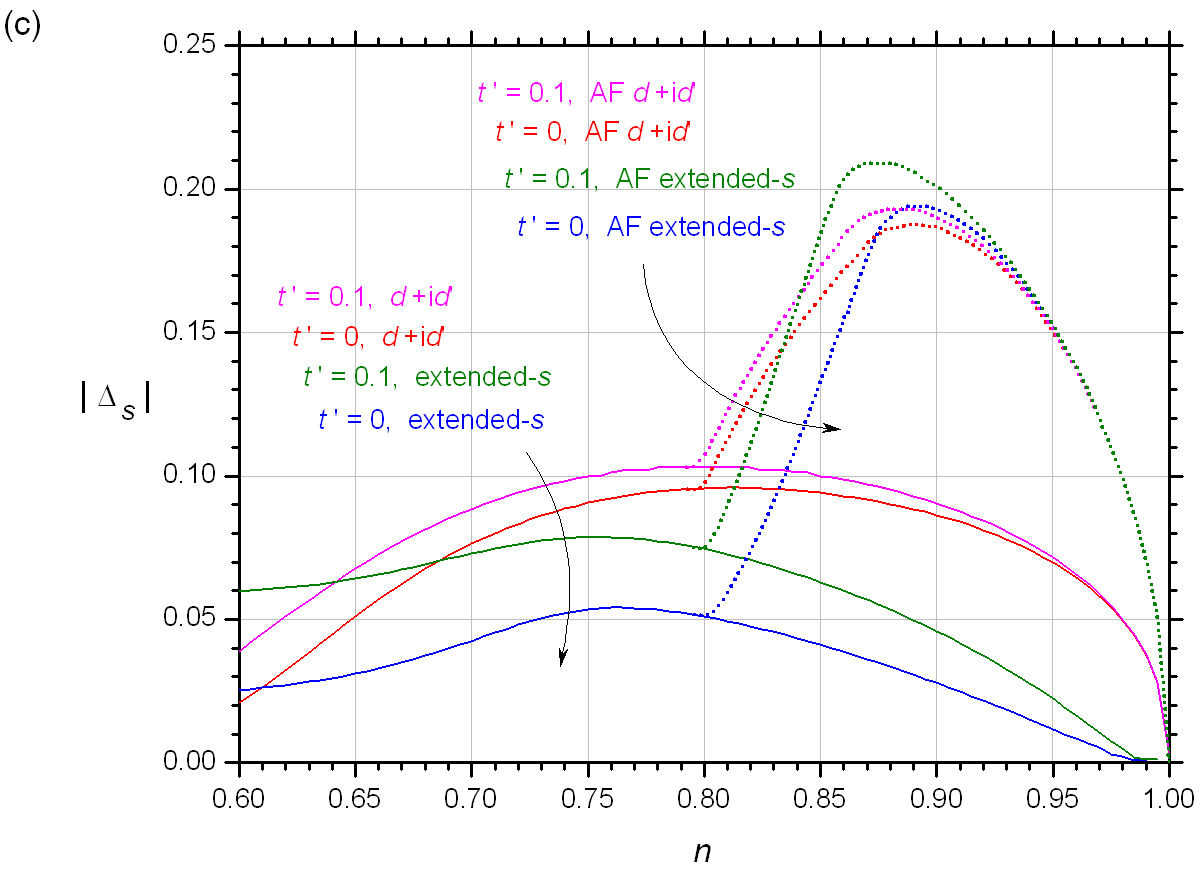} ~~~~~~~~~~~~~~~
\includegraphics  [scale=0.36]  {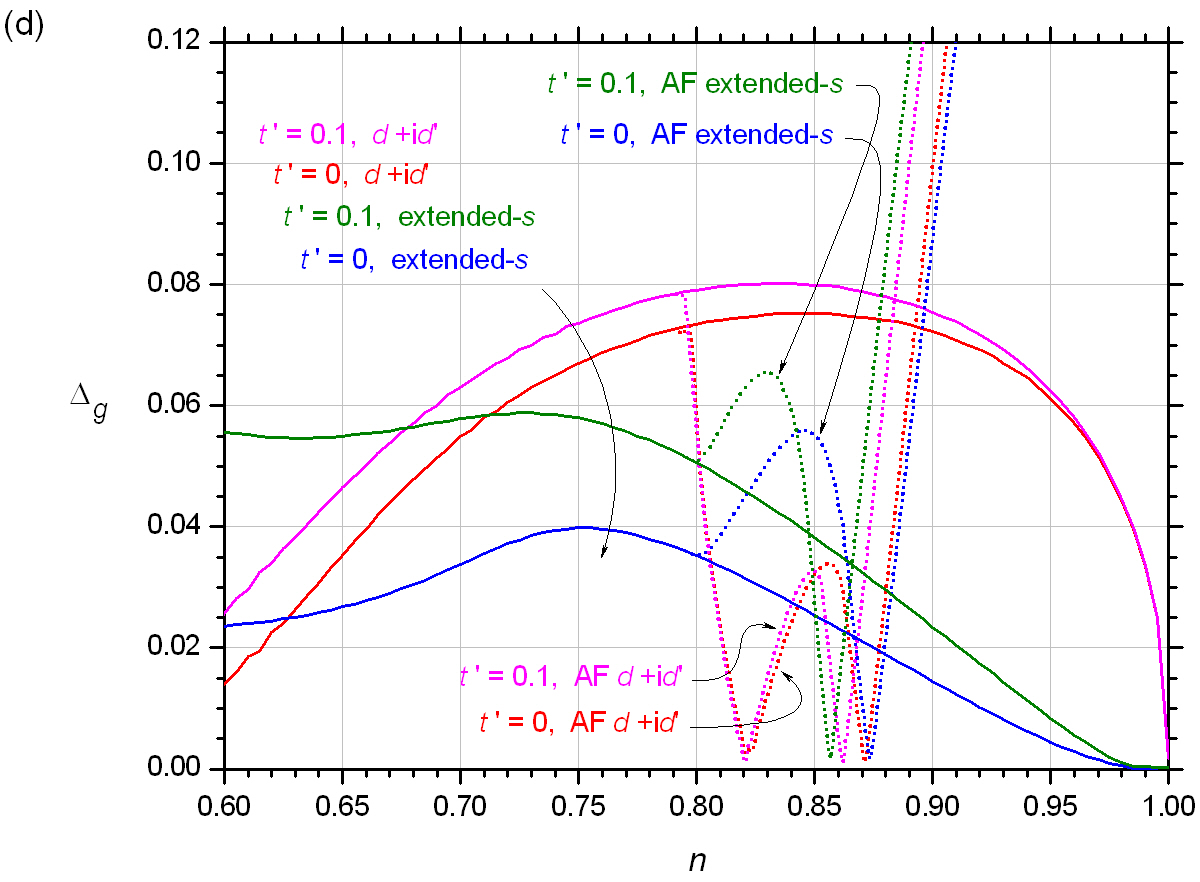} \\

\onehalfspacing

\caption  {
This figure plots the AF order $|m|$, superconducting order $|\Delta_s|$, and full energy gap $\Delta_g$ against the band filling fraction $n$,  in a number of nonmagnetic and AF, chiral-$d$ wave and extended-$s$ wave superconducting systems at zero temperature.
For a system with $(t,t';J) = (-1, 0 ; 0.5)$, we plot (a) the $|\Delta_s|$ and $|m|$ in different states, and (b) the $\Delta_g$ in different states.
For a system with $(t,t';J) = (-1,0;1)$ or $(-1,0.1;1)$, we plot (c) the $|\Delta_s|$ in different states, and (d) the $\Delta_g$ in different states.
}
\label  {xs-op}
\end  {figure}

In order to gain understanding on the dependence of single particle dispersion relation on the symmetry of the SC, and also to identify features pertinent to the AF chiral-$d$ wave SC, we make a comparison with the AF extended-$s$ wave SC. It is noted that previous studies show that in low doping systems, the extended-$s$ wave SC has a stability next to the chiral-$d$ wave SC \cite {Voo20}.
We also study the sensitivity of the results to the band structure and interaction strength.
Fig.~\ref {xs-op} compares the AF order $|m|$, superconducting order $|\Delta_s|$, and full energy gap $\Delta_g$  in the AF, nonmagnetic, chiral-$d$ wave, and extended-$s$ wave states. We consider only zero temperature systems.
Figs.~\ref {xs-op}(a) and \ref {xs-op}(b) consider the system parameters $(t,t';J) = (-1, 0 ; 0.5)$; while Figs.~\ref {xs-op}(c) and \ref {xs-op}(d) consider more strongly interacting systems, with and without a 2NN hopping, where $(t,t';J) = (-1, 0 ; 1)$ or $(-1, 0.1 ; 1)$.

% ex-s and chiral-d :  D_s  for  J = 0.5

For the system parameters $(t,t';J) = (-1, 0 ; 0.5)$, Fig.~\ref {xs-op}(a) plots the $|\Delta_s|$ and $|m|$ at zero temperature against the band filling fraction $n$.
There are a number of notable points in a comparison between the chiral-$d$ wave and extended-$s$ wave states.
The most of all is the enhancement in $|\Delta_s|$ by the AFM regardless of the SC symmetry.
The enhancement is even more dramatic in the extended-$s$ wave case. In a nonmagnetic environment, the $|\Delta_s|$ for an extended-$s$ wave SC is barely visible in the plot, when it is outside the range $0.65 \lesssim n \lesssim 0.85$. But when the system enters the AF regime, the $|\Delta_s|$ in the extended-$s$ wave SC is strongly enhanced, such that when it is in the very small doping density range $n \gtrsim 0.95$, the $|\Delta_s|$ in both SC symmetries become almost identical.
This indicates that the enhancement of the SC is is mainly due to the band structure. Presumably, the enhancement is due to the increased DOS at the FL, as a result of the opening of an AF gap above the FL.
The AFM is barely affected by the presence of SC, and it is also independent of the SC symmetry.

% ex-s and chiral-d :  D_g  for  J = 0.5

Fig.~\ref{xs-op}(b) plots the full energy gap $\Delta_g$ in those states discussed in Fig.~\ref{xs-op}(a). The $\Delta_g$ for the nonmagnetic extended-$s$ wave state is suppressed at $n \gtrsim 0.85$ simply due to the diminishing $|\Delta_s |$.
In the AF regime, the profiles of the $\Delta_g$ in extended-$s$ wave SC and chiral-$d$ wave SC are qualitatively different.
We see that nodal excitation in an AF extended-$s$ wave state can occur only at one band filling fraction, whereas a chiral-$d$ wave state can have a nodal excitation at two band filling fractions. These behaviors can be seen more clearly with a stronger $J$, and we will return to a more detailed discussion for this issue in Fig.~\ref {xs-op}(d).

% ex-s and chiral-d :  D_s  for  J = 1  with/without   t' = 0

For the two sets of system parameters $(t,t';J) = (-1,0;1)$ and $(-1,0.1;1)$, where the later includes a 2NN hopping,
Fig.~\ref{xs-op}(c) plots the zero temperature $|\Delta_s|$ in different states, including nonmagnetic, AF,  extended-$s$ wave, and chiral-$d$ wave states against the band filling fraction $n$. We dismiss the plot for $|m|$ since it always has a similar monotonic featureless profile almost independent of the presence and symmetry of the SC.
This figure is meant to examine the robustness of the results against the system parameters. A stronger Heisenberg exchange integral also amplifies the superconducting order for us to see details in the single particle dispersion relations more clearly.
The AF regime in this figure is expanded to $n \gtrsim 0.8$.
Focusing only on the qualitative features in the curves, we see that the $|\Delta_s|$'s in Figs.~\ref{xs-op}(a) and \ref{xs-op}(c) in the band filling fraction range $n \gtrsim 0.75$ are similar in the following aspects.
In nonmagnetic states, the $|\Delta_s|$'s in chiral-$d$ wave states are larger than those in extended-$s$ wave states.
In strongly AF states, there is a band filling fraction range where the $|\Delta_s|$'s in chiral-$d$ wave states and extended-$s$ wave states are comparable (see the curves at $n \gtrsim 0.95$ in Fig.~\ref{xs-op}(a); and $n \gtrsim 0.9$ in Fig.~\ref{xs-op}(c)).
We always observe these behaviors regardless of the variation in the band structure by a 2NN hopping integral.

% ex-s and chiral-d :  D_g  for  J = 1  with/without   t' = 0

Fig.~\ref{xs-op}(d) plots the full energy gap $\Delta_g$ for those states in Fig.~\ref{xs-op}(c). The strong $J$ brings out details in the $\Delta_g$'s in both SC symmetries clearly. Too, when we focus on only the qualitative features, the $\Delta_g$'s in Figs.~\ref{xs-op}(b) and \ref{xs-op}(d) are similar.
The AF chiral-$d$ wave states have vanishing $\Delta_g$ at two band filling fractions; while the AF extended-$s$ wave states have vanishing $\Delta_g$ only at one band filling fraction.
In Fig.~\ref{xs-op}(d), the vanishing $\Delta_g$ at $0.85 \lesssim n \lesssim 0.88$, where $\Delta_g$ rises monotonically with decreasing doping density at going beyond this vanishing point,  is common to both SC symmetries.
On the other hand, the vanishing $\Delta_g$ at $n \simeq 0.82$, occurs exceptionally in AF chiral-$d$ wave states.
Furthermore, the $\Delta_g$'s in the two SC symmetries respond oppositely to the onset of AFM at $n \simeq 0.79$. The $\Delta_g$ for a chiral-$d$ wave state sharply drops, but the $\Delta_g$ for an extended-$s$ wave state sharply rises at the onset of AFM.
These observations are also insensitive to the band structure and interaction strength.

% recapitulation features in |D_s| and D_g profiles in the two SCs.

A recapitulation of the above study on the $|\Delta_s|$ and $\Delta_g$ in chiral-$d$ wave and extended-$s$ wave states may be given below.
The profile of $|\Delta_s|$ does not qualitatively distinguish a chiral-$d$ wave state from an extended-$s$ wave state, regardless of the system is in a nonmagnetic or an AF state.
On the other hand, when the systems are AF, the profile of $\Delta_g$ exhibits a qualitative difference between the two SC symmetries. The AF chiral-$d$ wave states show vanishing $\Delta_g$ at two band filling fractions, whereas the AF extended-$s$ states show vanishing $\Delta_g$ only at one band filling fraction. These observations are robust against variation in the band structure and interaction strength.

% arrangement of the dispersion relation fig. panels

\begin {figure}

\includegraphics  [scale=0.4]  {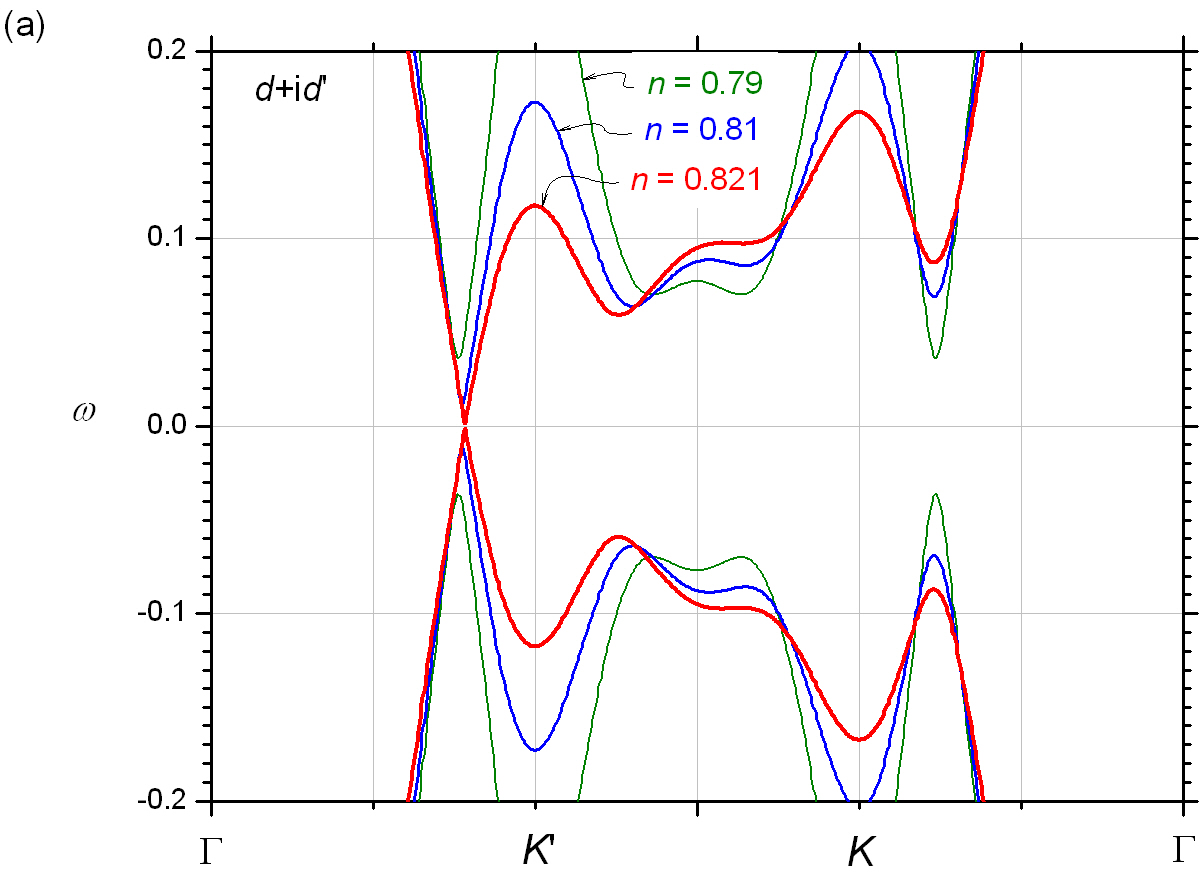} ~~~~~~~~
\includegraphics  [scale=0.4]  {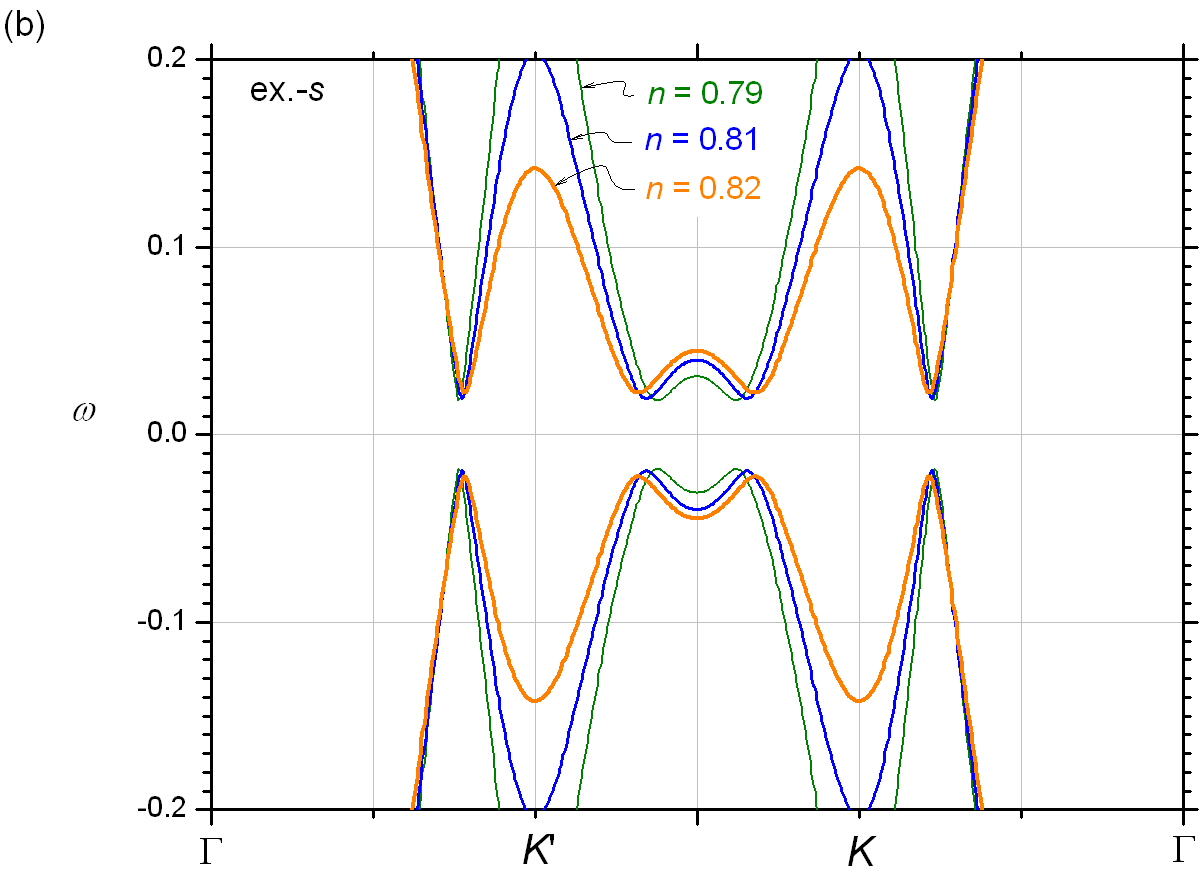} \\

\includegraphics  [scale=0.4]  {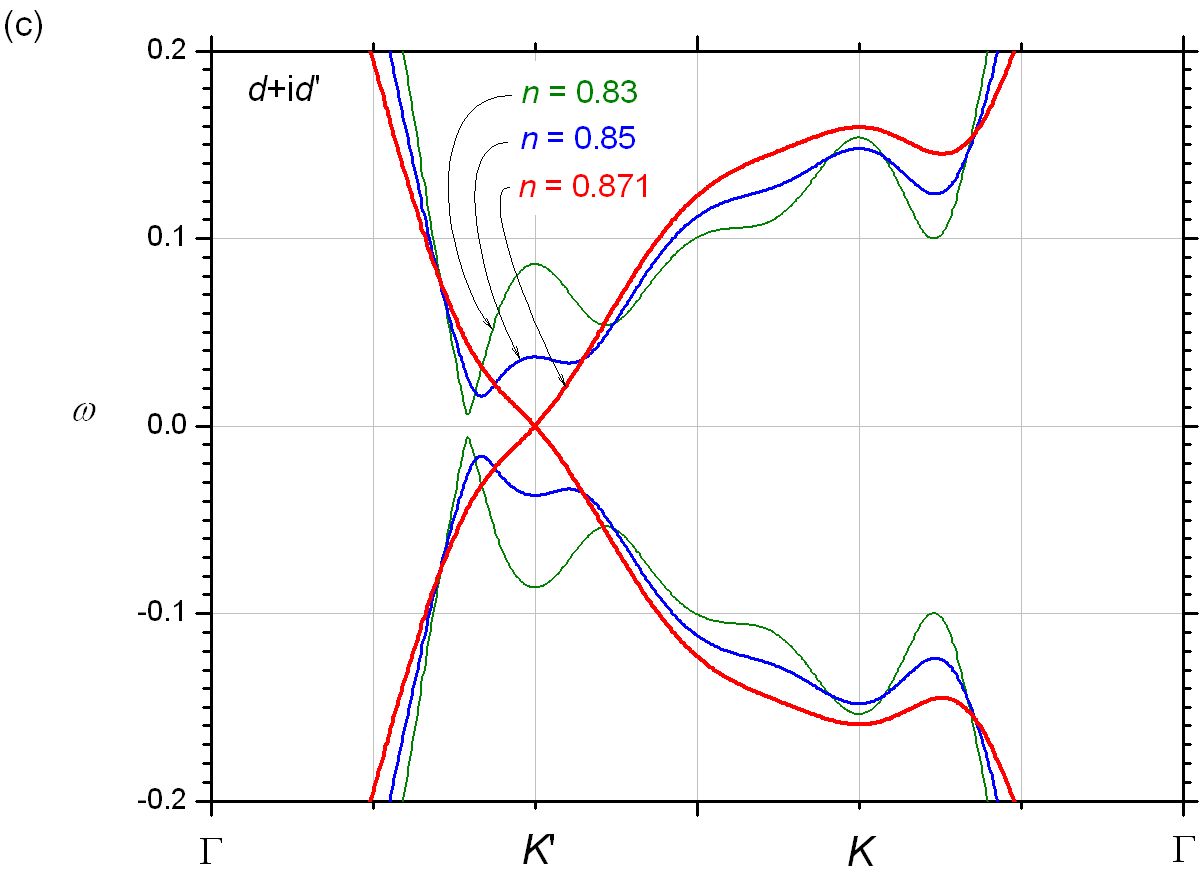} ~~~~~~~~
\includegraphics  [scale=0.4]  {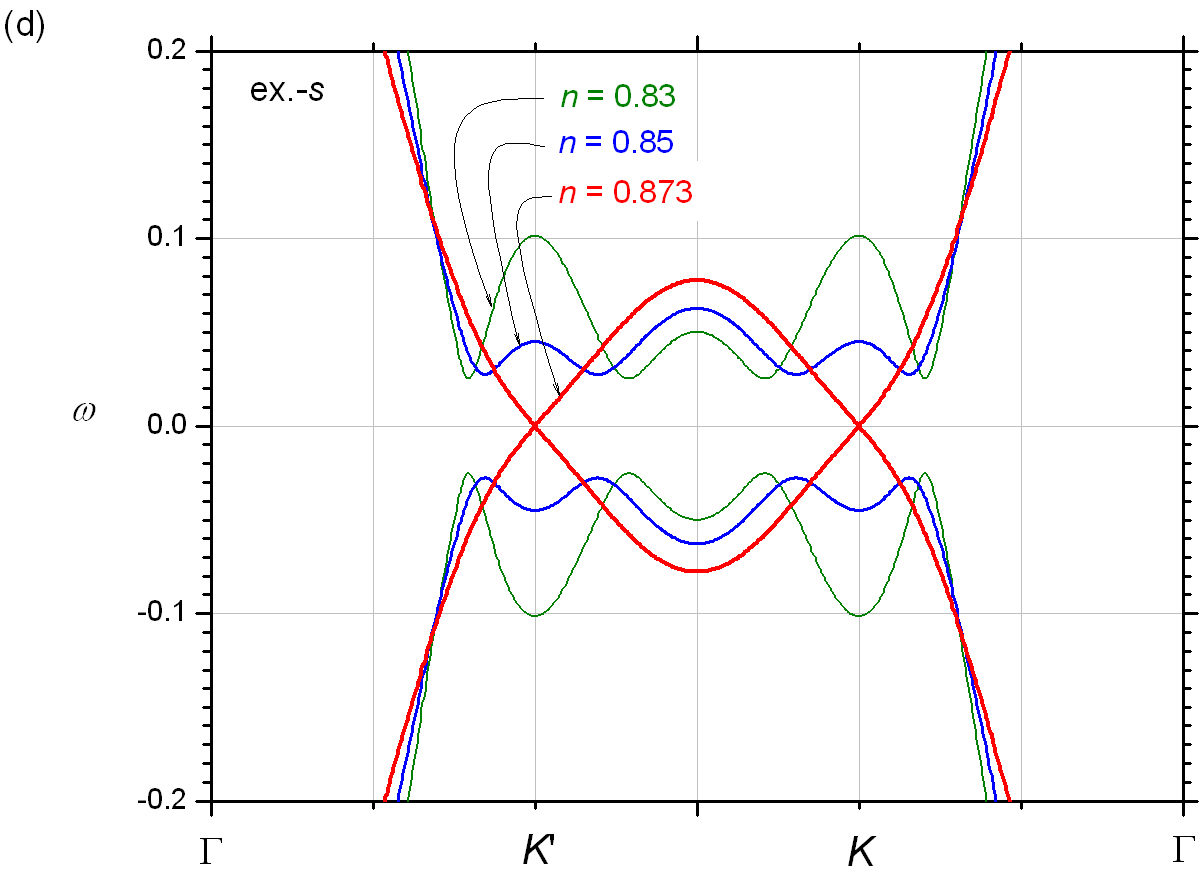} \\

\includegraphics  [scale=0.4]  {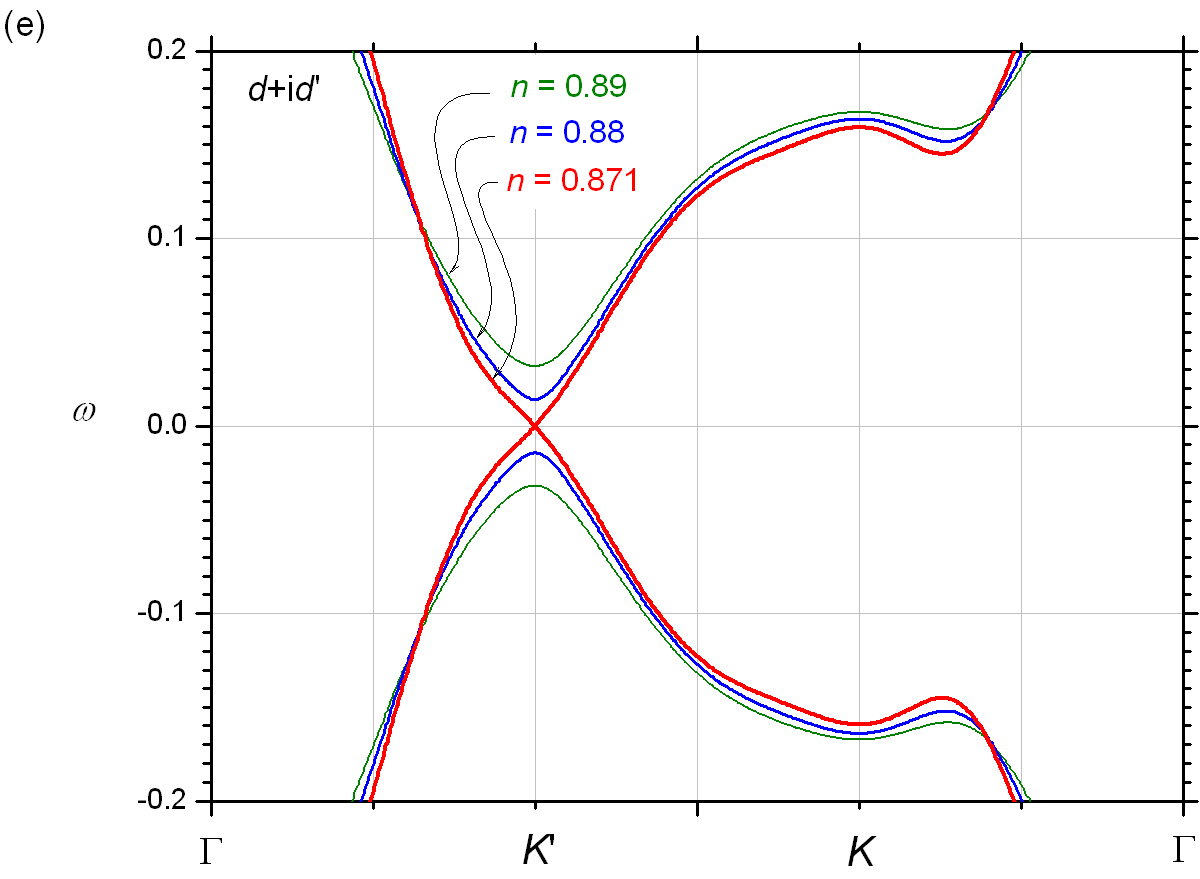} ~~~~~~~~
\includegraphics  [scale=0.4]  {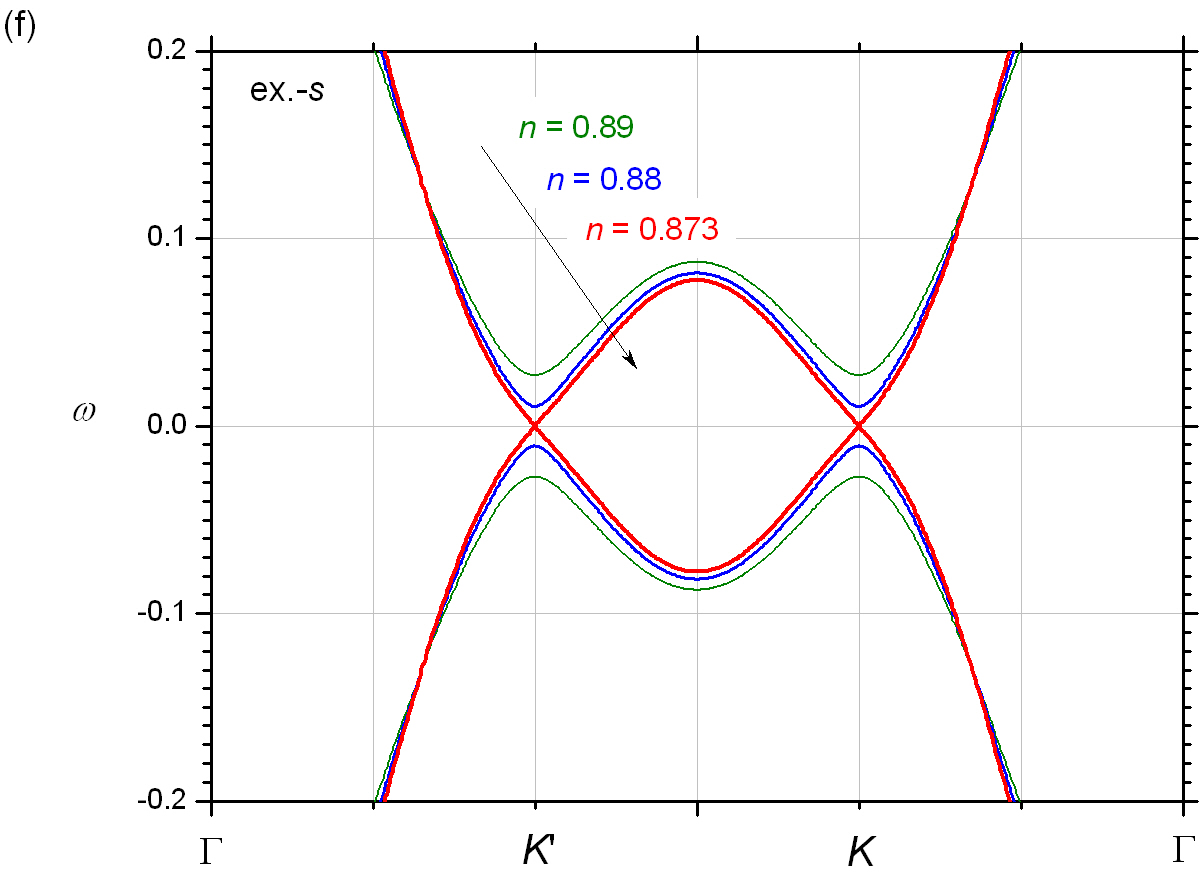} \\

\onehalfspacing

\caption  {
The low frequency part of the single particle dispersion relation $\varepsilon _{{\bf k, \uparrow}}$ is plotted along $\Gamma - K'- K - \Gamma$. We consider systems with $(t,t';J) = ( -1, 0 ; 1 )$ and $\langle \hat S ^{z, {\rm SB}} _{{\bf i}A} \rangle \geqslant 0$.
The SC symmetries and band filling fractions in the panels are
(a) $d+id'$ wave and $n=0.79$, 0.81, and 0.821;
(b) extended-$s$ wave and $n=0.79$, 0.81, and 0.82;
(c) $d+id'$ wave and $n=0.83$, 0.85, and 0.871;
(d) extended-$s$ wave and $n=0.83$, 0.85, and 0.873;
(e) $d+id'$ wave and $n=0.871$, 0.88, and 0.89; and
(f) extended-$s$ wave and $n=0.873$, 0.88, and 0.89.
}
\label  {xs-ep}
\end  {figure}

Fig.~\ref {xs-ep} compares the single particle dispersion relations in the AF $d+id'$ wave and AF extended-$s$ wave states, to highlight the unique features in the chiral-$d$ wave dispersion relations.
We consider zero temperature systems with $(t,t';J) = ( -1,0; 1 )$ and $\langle \hat S ^{z, {\rm SB}} _{{\bf i}A} \rangle \geqslant 0$, and we plot the low frequency part of $\varepsilon _{{\bf k, \uparrow}}$ along the route $\Gamma - K'- K - \Gamma$.
The panels in Fig.~\ref {xs-ep} are arranged in the following manner.
Figs.~\ref {xs-ep}(a) and \ref {xs-ep}(c), and \ref {xs-ep}(e) plot the $\varepsilon _{{\bf k, \uparrow}}$ for $d+id'$ wave states ($\varphi _{\rm sc} = 2 \pi / 3$); and Figs.~\ref {xs-ep}(b) and \ref {xs-ep}(d), and \ref {xs-ep}(f) plot the $\varepsilon _{{\bf k, \uparrow}}$ for extended-$s$ wave states ($\varphi _{\rm sc} =  0$).
For the convenience in our discussion, we divide the AF regime into three subsections, and each panel plots the dispersion relations at a few selected band filling fractions belonging to a subsection.
Figs.~\ref {xs-ep}(a) and Fig.~\ref {xs-ep}(b) are plotted at the emergence of AFM, where $0.80 \lesssim n \lesssim 0.83$;
Figs.~\ref {xs-ep}(c) and \ref {xs-ep}(d) are plotted at the intermediate AF strength regime, where $0.83 \lesssim n \lesssim 0.87$;
and Figs.~\ref {xs-ep}(e) and \ref {xs-ep}(f) are plotted in the strongly AF regime, where $0.87 \lesssim n < 1$.
The $|\Delta_s|$'s and $\Delta_g$'s for the above considered systems have been given in Figs.~\ref {xs-op}(c) and \ref {xs-op}(d).

% symmetries in ex-s states :  symm. K' and K.  isotropic at K'/K

The single particle dispersion relations in the extended-$s$ wave states possess a symmetry in $K'$ and $K$, throughout the nonmagnetic and AF regimes.
We see that the dispersion relations along $K' - \Gamma$ and $K - \Gamma$ are identical in all cases (Figs.~\ref {xs-ep}(b), \ref {xs-ep}(d),  and \ref {xs-ep}(f)). In the chiral-$d$ wave states, we can only see this symmetry in nonmagnetic states, e.g., in the nonmagnetic chiral-$d$ wave state at $n = 0.79$ (Fig.~\ref {xs-ep}(a)).
When an AFM rises in a chiral-$d$ wave state, it affects the dispersion relation asymmetrically at $K'$ and $K'$ (Figs.~\ref {xs-ep}(a), \ref {xs-ep}(c),  and \ref {xs-ep}(e)).
Furthermore, the dispersion relations for extended-$s$ wave states may be considered as more isotropic at $K'$ and $K$.
In the extended-$s$ wave states, we see that all the extrema of the dispersion relations on the route $\Gamma - K' - K - \Gamma$ are at the same level, throughout the nonmagnetic and AF regimes (Figs.~\ref {xs-ep}(b), \ref {xs-ep}(d), and \ref {xs-ep}(f)).
The chiral-$d$ wave states do not possess such a symmetry, even when they are nonmagnetic. For example, in the  nonmagnetic chiral-$d$ wave dispersion relation at $n = 0.79$, the extrema in between $K'$ and $K$ are at a higher frequency, compared with those extrema at $K'-\Gamma$ and $K - \Gamma$ (Fig.~\ref {xs-ep}(a))

% emerging AFM regime

Figs.~\ref {xs-ep}(a) and \ref {xs-ep}(b) show the evolution of the dispersion relation $\varepsilon _{{\bf k, \uparrow}}$ in an emerging AFM, for $d+id'$ wave and extended-$s$ wave states respectively.
In the $d+id'$ wave states, the AF effect is asymmetric at $K'$ and $K$, where the dispersion near $K'$ is softened, eventually results in nodes near $K'$ at $n = 0.821$. The corresponding $\Delta_g$ depends sensitively on the band filling fraction.
In the extended-$s$ wave states, the $\Delta_g$ is rather insensitive to the change of the band filling fraction.
The nodal state at $n = 0.821$ is an unique feature in the $d+id'$ wave states. The state comprises a triplet of dispersion nodes encircling $K'$ (see a similar example in Fig.~\ref {aksw_zoom}(e)).

% intermediate afm regime

Figs.~\ref {xs-ep}(c) and \ref {xs-ep}(d) show the dispersion relation $\varepsilon _{{\bf k, \uparrow}}$ in the intermediate AF strength regime, for $d+id'$ wave and extended-$s$ wave states respectively.
As the systems approach HBF, the low frequency dispersion extrema congregate and coalesce, and a coalescence is accompanied by the formation of a dispersion node with a pair of upward and downward touching Dirac cones. For example, the low frequency extrema in the $d+id'$ wave states congregate into and merge at $K'$, at $n = 0.871$; and the low frequency extrema in the extended-$s$ wave states congregate into and merge at both $K'$ and $K$, at $n = 0.873$.
We note that the congregating shift of the dispersion extrema actually has quietly started since the emergence of the AFM (refer to Figs.~\ref {xs-ep}(a) and \ref {xs-ep}(b)).
The $\Delta_g$'s in both superconducting symmetries are sensitive to the band filling fraction in the intermediate AF strength regime.

% origin of AF ex-s gap node at K' and K

An analytic analysis of the dispersion node at $K'$ in an AF $d+id'$ wave state has been presented for a similar case in Fig.~\ref {aksw_zoom}(b).  A parallel analysis can be presented for the dispersion nodes at $K'$ and $K$ in an AF extended-$s$ wave state (Fig.~\ref {xs-ep}(d)).
In an extended-$s$ wave state,  $\varphi _{\rm sc} = 0$ leads to the same profile for $| \Delta^{AB} _{{\bf k} , \sigma } |$, $| \Delta ^{AB} _{- {\bf k} , { \bar \sigma } } |$, and $| \varepsilon^{AB} _{\bf k} |$, which vanishes at both $K'$ and $K$.
This reduces $D_{{\bf k},\sigma}$ to $D_{{\bf k},\sigma} =
( {\varepsilon ' _{ {\bf k} A \sigma } } {\varepsilon ' _{{ \bf k} A \bar \sigma } } ) ^ 2$ at both $K'$ and $K$ (note that this $D_{{\bf k},\sigma}$ is independent of the sign of $\eta_\sigma m$), which vanishes at some point in the course of a growing $|m|$ at approaching HBF.

% strong AF regime

Figs.~\ref {xs-ep}(e) and \ref {xs-ep}(f) show the evolution of the dispersion relation $\varepsilon _{{\bf k, \uparrow}}$ in the strong AFM regime, for $d+id'$ wave and extended-$s$ wave states respectively.
The evolution of the dispersion relations in the change of band filling fraction is rather simple. The tips of the Dirac cones are rounded off, the dispersion nodes are lifted, and the $\Delta_g$'s in both symmetries of SC rapidly grow as the systems approaches HBF.
The observation that the vanishing of the $\Delta_g$'s at $n \simeq 0.87$ (Fig.~\ref {xs-op}(d)) occurs near the maxima of the $|\Delta_s|$'s (Fig.~\ref {xs-op}(c)) may seem counterintuitive.
Even more counterintuitive, when we go beyond this band filling fraction,  we see growing $\Delta_g$'s but diminishing $|\Delta_s|$'s at approaching HBF.
The shrinkage of the FS at approaching HBF removes the habitat of the SC, and inevitably results in a vanishing $\Delta_s$ when one goes beyond certain band filling fraction.
But the dispersion gap $\Delta_g$ is an intertwined result of the AFM and SC. In very low doping density states, $\Delta_g$ has more an AF nature, and its magnitude reflects the strength of the AFM.

% probe of chiral-d and ex-s symmetry in AF states

Through Figs.~\ref {aksw_zoom}, \ref{xs-op}, and \ref {xs-ep}, we may sort out some unique features, which are robust against the band and interaction parameters, in the AF chiral-$d$ wave states.
When a system is nonmagnetic, we may not readily identify a chiral-$d$ wave state from an extended-$s$ wave state with a non-phase-sensitive probe. But when there is a coexisting AFM in the system, the asymmetric effect on the single particle dispersions in the chiral-$d$ wave states provides an alternative mean to the problem. One may perform a spin-sensitive measurement on the electrons ejected from the lattice with angle resolved photoemission spectroscopy to identify the superconducting phase.
Moreover, we can observe nodally excited AF chiral-$d$ wave states at two band filling fractions.

\section  {Concluding remarks}
\label  {conc}

% recapitulate the findings

We have studied the AFM and SC in a honeycomb lattice that is lightly doped from a Mott state, with a $t-J$ model and a SB MFT. We focus on the AF chiral-$d$ wave states, and compare it with the AF extended-$s$ wave states.  We find that there is always an enhancement in the SC by the AFM regardless of the symmetries of the SC.
The single particle dispersion relation $\varepsilon _{{\bf k, \sigma}}$ in an AF chiral-$d$ wave state depends on the sign of $\varphi_{\rm sc} \eta_\sigma m$, and is asymmetric in $K'$ and $K$. The $\varepsilon _{{\bf k, \sigma}}$ in an AF extended-$s$ wave state is symmetric in $K'$ and $K$.
We can always find nodally gapped AF superconducting states.
We find nodally gapped AF chiral-$d$ wave states at two band filling fractions.
On the other hand,  we can only find a nodally gapped AF extended-$s$ wave state at one band filling fraction.
All single particle dispersion nodes are in the form of a pair of Dirac cones touching each other like the Minkowski light cone in the relativity theory.

% comparing honeycomb and square lattices

It is interesting to make a comparison between the low doping $t-J$ models in two-dimensional honeycomb lattice and square lattice.
The first to note is the contrasting phenomena where an AFM enhances the SC in a honeycomb lattice, but suppresses the SC in a square lattice \cite {Voo17}.
The strength of a SC in a system may be understood by studying the DOS and the FS.
An AFM always enhances the DOS near the FL, in both the honeycomb lattice and square lattice, due to the accumulation of the states expeled from the AF gap. But an AFM affects the FSs in the two lattices in very different ways.
Before the discussion of the FSs in the two lattices, it is reminded that the preferred superconducting order symmetries in low doping honeycomb and square lattices are the nodeless chiral-$d$ wave and nodal $d_{x^2-y^2}$ wave respectively.
A honeycomb lattice is intrinsically a two-sublattice structure, and it can accommodate an AFM with no significant change in the FS. Whereas a square lattice is a one-sublattice structure, where an AFM breaks the translational symmetry and destroys the FS at the magnetic Brillouin zone boundary. Since the antinodes of the $d_{x^2-y^2}$ wave are located near the destructed sections of the FS, the habitate of the SC is deprived.
Therefore an AFM enhances the SC in a honeycomb lattice simply by enhancing the DOS; whereas an AFM suppresses the SC in a square lattice due to a destruction of the FS.
There is another contrasting finding between the two lattices. In a  honeycomb lattice,  though the chiral-$d$ wave order is nodeless, an AF chiral-$d$ wave state can be nodally gapped; whereas in a square lattice, though the $d_{x^2-y^2}$ wave order is nodal, an AF $d_{x^2-y^2}$ wave state can be nodeless gapped, since the $d_{x^2-y^2}$ wave nodes on the FS can be destroyed by the AFM \cite {VW05}.

% closing

We have discussed the order parameters and single particle dispersion relations in the AF chiral-$d$ wave superconducting states in detailed. Though the chiral-$d$ wave and extended-$s$ wave states may show only minor difference in a nonmagnetic environment, they exhibit distinctive behavior when there is a coexisting AFM in the system. This study may be related to the AF and insulating half-filled honeycomb lattice material In$_3$Cu$_2$VO$_9$, which might be a Mott material.

\noindent {\bf Acknowledgment} This work is not supported by any organization.

%.....................................................................................
%....................................................................................

%.....................................................................................
% ...
%....................................................................................

%.....................................................................................
% ...
%....................................................................................

%.....................................................................................
% identical symbol size scales are used in all panels.
%....................................................................................

%.....................................................................................
% ...
%....................................................................................

%.....................................................................................
% ...
%....................................................................................

%.....................................................................................
% ...
%....................................................................................

\end {document}